\documentclass[12pt,reqno]{amsart}                  % delete ",reqno" to put equation numbers on left

\usepackage[dvips]{graphicx}                        % I think this is for using EPS files, but not sure
\usepackage{latexsym}                               % not sure what this is
\usepackage{amssymb}                                % not sure what this is

\newcounter{maintheoremnumber}
\newtheorem{theorem}{Theorem}[section]                % delete "[section]" to number consecutively
\newtheorem{mainthm}[maintheoremnumber]{Theorem}
\newtheorem{lemma}{Lemma}[section]

\newtheorem{proposition}[lemma]{Proposition}

\renewcommand{\epsilon}{\varepsilon}

\newcommand{\bth}{\begin{theorem}}
\newcommand{\n}{\noindent}
\newcommand{\dis}{\displaystyle}
\newcommand{\re}{\, \textup{Re} \,}
\newcommand{\im}{\, \textup{Im} \,}
\newcommand{\R}{\mathbb{R}}
\newcommand{\C}{\mathbb{C}}
\newcommand{\ba}{\begin{align*}}
\newcommand{\ea}{\end{align*}}
\usepackage[colorlinks=true, hypertex, linkcolor=blue, urlcolor=blue, citecolor=blue]{hyperref}
\title{Density of Complex Zeros of a System of Real Random Polynomials}
\author{Brian Macdonald}
\email{bmac@jhu.edu}
\thanks{Thanks to Bernard Shiffman, for his guidance, teaching, patience, and
for countless helpful meetings, conversations and suggestions.}
\date{\today}
\begin{document}
%\begin{center} NOT FOR CIRCULATION \end{center}
\begin{abstract}
We use probability and several complex variable theory to study the density of complex zeros of a system of real random polynomials in several variables.
We use the Poincar\'{e}-Lelong formula to show that the density of complex zeros of a random polynomial system with real coefficients rapidly approaches the density of complex zeros in the complex coefficients case.  We also show that the behavior the scaled density of complex zeros near $\mathbb{R}^m$ of the system of real random polynomials is different in the $m\geq 2$ case than in the $m=1$ case: the density goes to infinity instead of tending linearly to zero.
\end{abstract}
\maketitle
\tableofcontents
\pagenumbering{arabic}
%%%%%%%%%%%%%%%%%%%%%%%%%%%%%%%%%%%%%%%%%%%%%%%%%%%%%%%%%%%%%%%%%%%%%%%%%%%%%
%\input{Intro-to-Zeros}
\section{Introduction}
  The density of real (resp. complex) zeros of random polynomials in one and several variables with real (resp. complex) Gaussian coefficients has been studied by many.  Kac \cite{kac} and Rice \cite{rice} independently found the density of zeros of a random polynomial with real standard Gaussian coefficients.  Bogomolny, Bohigas, and Leboeuf (\cite{bbl92}, \cite{bbl96}) and Hannay \cite{hannay} have results on the density of (and correlations between) zeros of random polynomials with complex Gaussian coefficients.  Edelman and Kostlan \cite{ek} generalize the results for density of real (resp. complex) zeros to systems of independent random functions in several variables when the coefficients are real (resp. complex) Gaussian random variables.

   In one variable, Shepp and Vanderbei \cite{sv95}, Ibragimov and Zeitouni \cite{zeitouni}, and Prosen \cite{prosen} have studied \textit{complex} zeros of \textit{real} polynomials.  Shepp and Vanderbei extended Kac's formula for the density of zeros of polynomials in one real variable, in the case where the coefficients are standard real Gaussian coefficients, to include both real and non-real zeros of those same polynomials.  Ibragimov and Zeitouni studied the density of zeros of random polynomials with i.i.d coefficients (which are not necessarily Gaussian).  Prosen followed Hannay's approach and found both an unscaled and a scaled density formula for the complex zeros of a random polynomial with independent real Gaussian coefficients.

\begin{figure}[b]
\begin{center}

\includegraphics[width=0.5\textwidth]{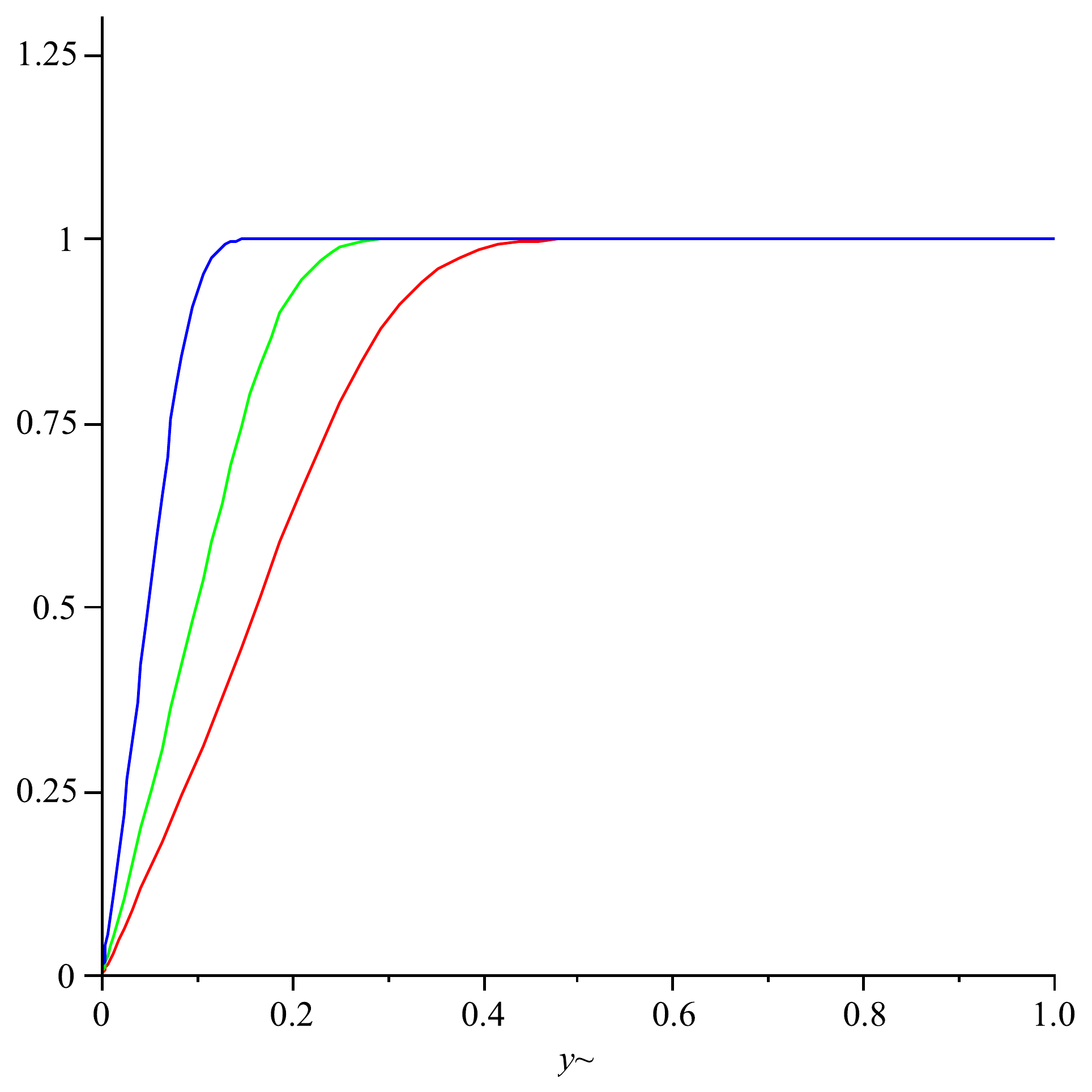}
%\includegraphics[width=0.5\textwidth]{scaledzerosinonevariableforthesis.eps}

%\put(-80,80){$\mathbf{{\gamma}_1}$}
\caption{The density of complex zeros of a random SO(2) polynomial for increasing values of $N$.  Because of symmetries, it is sufficient to plot the density along the imaginary axis for $0<y\leq 1.$ Here we have normalized so that the density of zeros of a random SU($2$) polynomial is the constant function 1.}\label{zeros1}

%\put(-80,80){$\mathbf{{\gamma}_1}$}
%\caption{Scaled density of zeros in two variables}
\end{center}
\end{figure}

   One consequence of Prosen's unscaled density formula is that, away from the real line, the density of complex zeros of an SO(2) random polynomial (which is the polynomial given by
   $$ f_N(z) = \sum_{j=0}^N a_j {N \choose j}^{\frac{1}{2}} z^j, $$
   where $a_j$ is a \textit{real} standard Gaussian random variable), rapidly approaches the density of complex zeros of a random SU(2) polynomial (which is the polynomial given by
   $$ f_N(z) = \sum_{j=0}^N c_j {N \choose j}^{\frac{1}{2}} z^j $$ where $c_j$ is a \textit{complex} standard Gaussian random variable), as $N$, the degree of the polynomial, goes to infinity. Figure \ref{zeros1} illustrates this fact.

\begin{figure}[b]
\begin{center}
\includegraphics[width=0.5\textwidth]{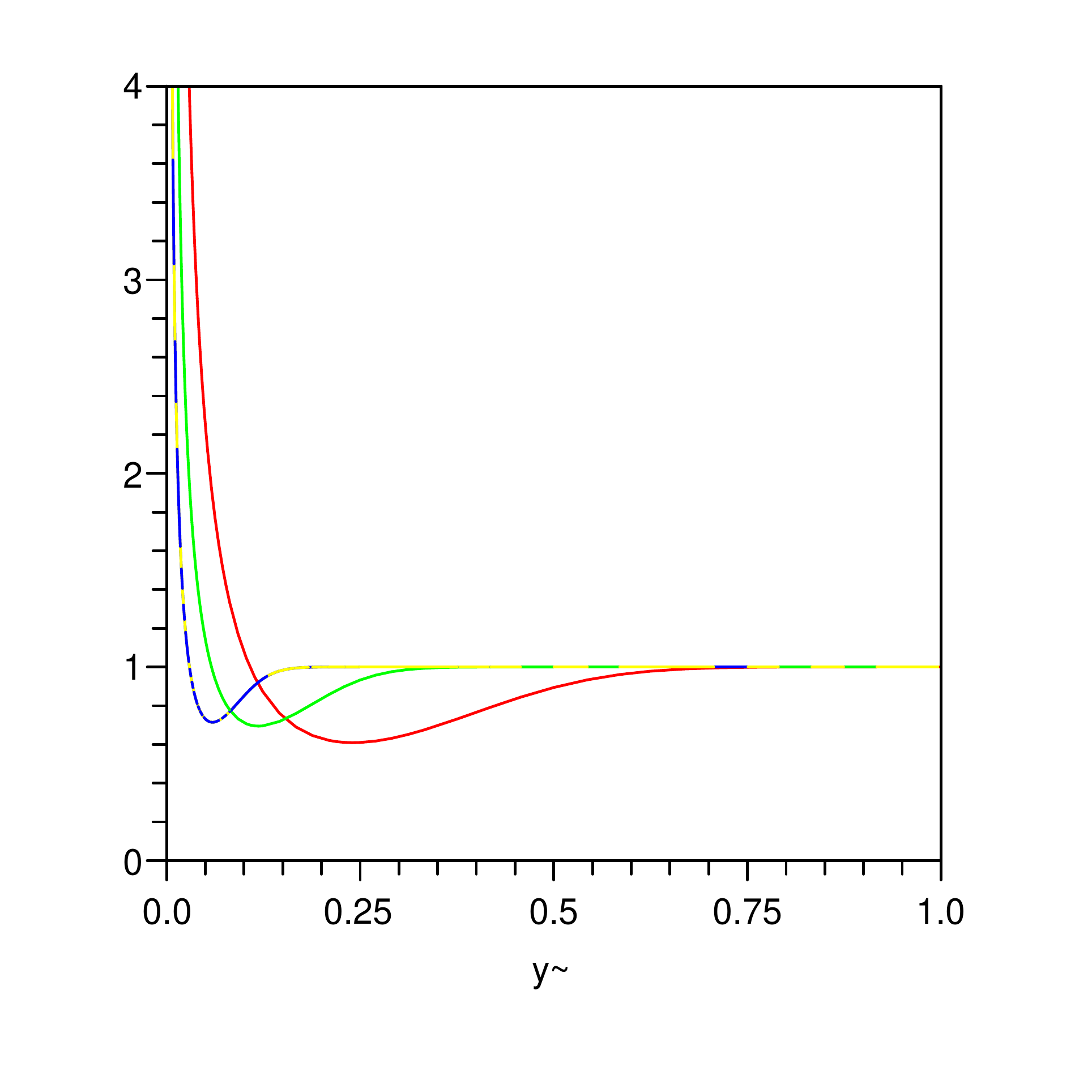}
%\put(-80,80){$\mathbf{{\gamma}_1}$}
\caption{The density of complex zeros of the SO($m+1$) polynomial for increasing values of $N$, in the $m=2$ case. Because of the invariance properties of the SO($m+1$) polynomial, it is sufficient to plot the density for $z =(iy,0)\in \mathbb{C}^m$ with $0<y\leq 1.$  We have normalized so that the density of zeros of the random SU($m+1$) polynomial is the constant function 1.}\label{zerosm}
\end{center}
\end{figure}

    Using the Poincar\'{e}-Lelong formula, we show this convergence, recovering Prosen's single variable result \cite{prosen}, and we show the convergence to be exponential.  In Theorem \ref{thm1} we generalize this result to the density of zeros of a random SO$(m+1)$ polynomial system in $m$ variables (defined below).  This generalized result is illustrated in Figure \ref{zerosm} for the special case $m=2.$  In \cite{bmcrit}, the author uses a different method  to generalize this result further to the density of critical points of a random SO($m+1$) polynomial in $m$ variables.

In one variable, Prosen also showed that, for every $N$, the density of zeros tends linearly towards zero as we approach the real line. This result can be seen in Figure \ref{zeros1}.  In several variables, the behavior near $||y||=0$ (i.e., near $\R^m$) is much different.  See Figure \ref{zerosm}.  The same can be said for the scaled density.  In Theorem \ref{thm2}, we show that near $\mathbb{R}^m$, the behavior of the scaled density of complex zeros of the system in Theorem \ref{thm1} is different in the $m\geq 2$ case than the behavior of the scaled density in the $m=1$ case that was shown by Prosen: for $m\geq 2,$ the density goes to infinity instead of tending linearly to zero.
\begin{figure}[b]
\begin{center}
\includegraphics[width=0.40\textwidth]{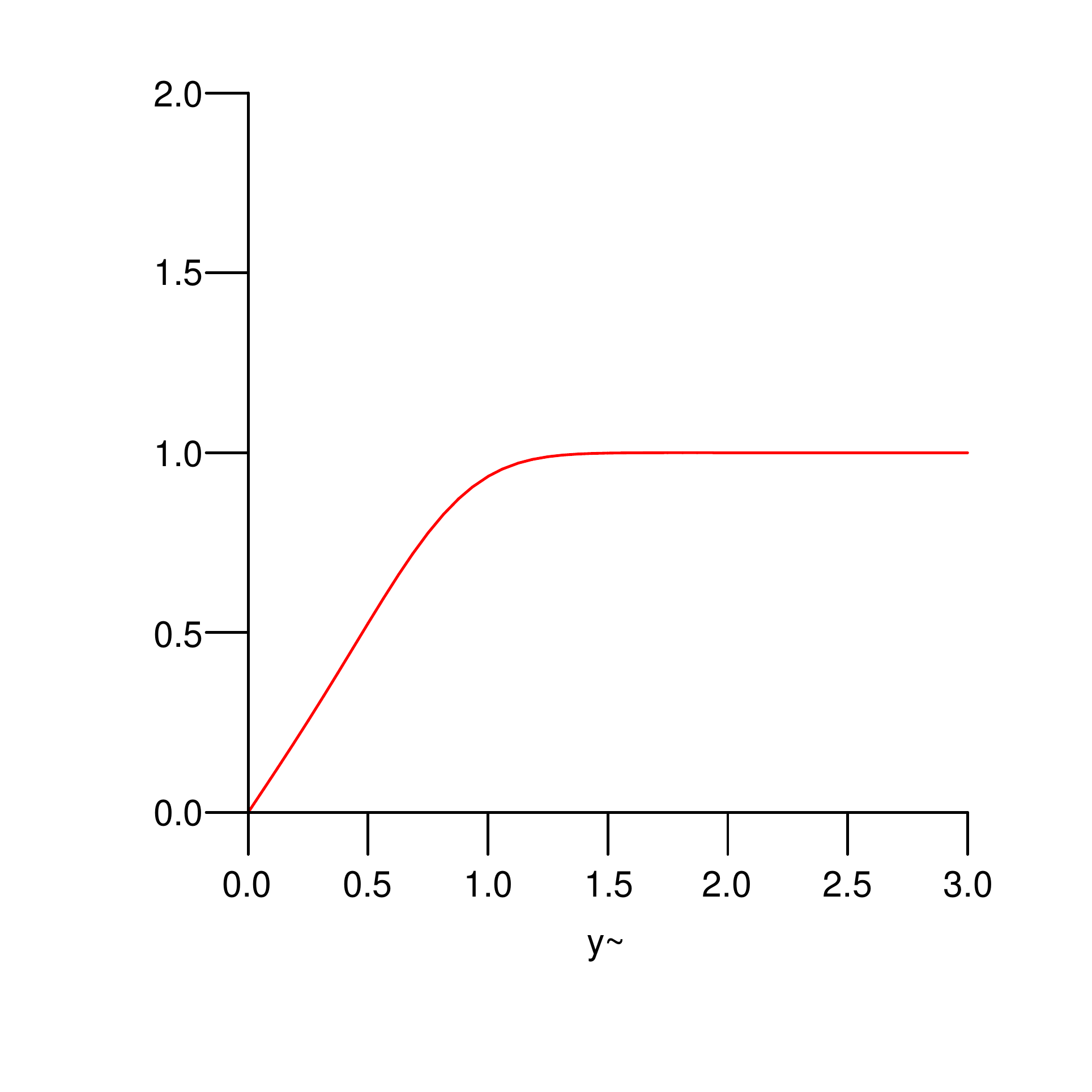}
\includegraphics[width=0.40\textwidth]{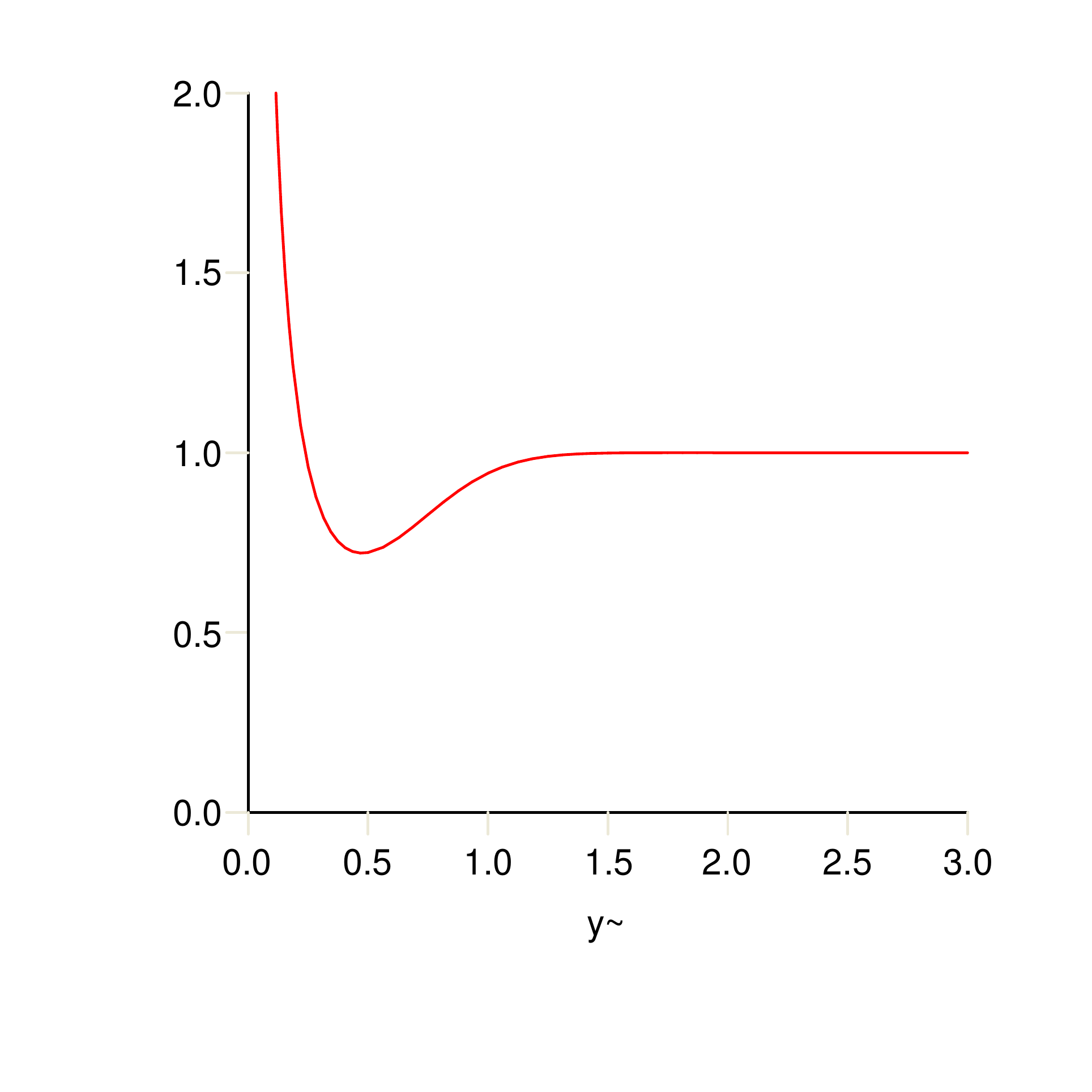}
%\put(-80,80){$\mathbf{{\gamma}_1}$}
\vskip -1cm
\caption{The scaled density of zeros for $m=1$ (left) and $m=2$ (right).}\label{scaledzeros}
%\put(-80,80){$\mathbf{{\gamma}_1}$}
%\caption{Scaled density of zeros in two variables}
\end{center}
\end{figure}
Figure \ref{scaledzeros} illustrates the scaled density for the SO($m+1$) polynomial for $m=1$ and $m=2.$

\subsection{Density of zeros} Consider $h _{m,N}=(f_{1,N}, ...\, , f _{m,N}):\mathbb{C}^m \rightarrow \mathbb{C}^m $, where
$f _{q,N}$ is an SO($m+1$) polynomial of the form
$$ f_{q,N}(z) = \sum_{|J|=0}^{N} c_J^q {N\choose J} ^{1/2} z^J,$$ where the $c^q_J$'s are independent complex random variables, where the random vector $\{c^q_J\}\in \mathbb{C}^{D_N}, D_N = {N+m \choose m}$ has associated measure $d\gamma$, and where we are using standard multi-index notation.  We show that for these $m$ independent functions in $m$ variables, the density of complex zeros in the real coefficients case rapidly approaches the density in the complex coefficients case as the degree of the polynomials gets large.  In fact, we show that the convergence is exponential.  Figure \ref{zerosm} illustrates this convergence for the case $m=2$.

Note that we have normalized these density functions so that the density in the complex coefficients case is the constant function 1, and that, because of invariance properties of the SO($m+1$) polynomial, it is enough to show the graph of the density function for $z=(iy,0) \in\C^m, 0 < y\leq 1.$

More formally, let
\begin{align*}
d\gamma_{cx} &=
\frac{1}{\pi^{N}}e^{-|c|^2} dc   \\
d\gamma_{real} &= \delta_{\mathbb{R}^{D_N}} \frac{1}{(2\pi)^{N/2}}e^{-|c|^2/2} dc,\end{align*}
 where $  c \in \mathbb{C}^{D_N},$ and $\delta_{\mathbb{R}^{D_N}}$ is the delta measure on ${\mathbb{R}^{D_N}} \subset \mathbb{C}^{D_N}.$  Here $d\gamma_{cx}$ corresponds to the standard complex Gaussian coefficients case, where we are considering the random SU($m+1$) polynomial
$$f_{q,N}(z) = \sum_{|J|=0}^{N} c^q_J {N\choose J} ^{1/2} z^J, $$
where the $c^q_J$'s are standard complex Gaussian random variables, and $d\gamma_{real}$ corresponds to the standard real Gaussian coefficients case, where we have the random SO($m+1$) polynomial
$$ f^{}_{q,N}(z) = \sum_{|J|=0}^{N} c^q_J {N\choose J} ^{1/2} z^J = \sum_{|J|=0}^{N} a^q_J {N\choose J} ^{1/2} z^J,$$
where $c^q_J = a^q_J +i0$ is a standard real Gaussian random variable.  Let $E_\gamma(\cdot)$ denote the expectation with respect to $\gamma$; or, in other words, integration over $\C^{D_N}$ with respect to $d\gamma.$  Let $\dis Z_{h_{m,N}(z)} = \sum_{h_{m,N}(z)=0} \kern -1em \delta_z\,\,$   denote the distribution corresponding to the zeros of $h_{m,N}(z).$  Then $E_\gamma(Z_{h_{m,N}(z)})$ denotes the density of the zeros of $h$ with respect to the measure $d\gamma.$  We now formally state the first result:
\begin{mainthm}\label{thm1}\addcontentsline{toc}{subsection}{Theorem \ref{thm1}}
 \begin{align*} \dis E_{\gamma_{real}}(Z_{h_{m,N}(z)} ) &=  E_{\gamma_{cx}}(Z_{h_{m,N}(z)} ) +O(e^{-\lambda_z N}),
\end{align*}
for all $z \in \mathbb{C}^m \backslash \mathbb{R}^m,$ where $\lambda_z$ is a positive
constant that depends continuously on $z$.  The explicit formula for $\lambda_z$ is
$$ \lambda_z = - \log\Big|\frac{1+z\cdot z}{1+||z||^2}\Big|. $$
Also, for compact sets $K \subset \mathbb{C}^m\backslash\mathbb{R}^m$, the density converges uniformly with an error term of  $O(e^{-\lambda_K N})$, where $\lambda_K$ is a constant that depends only on $K$.
\end{mainthm}

Note that for $z \in \mathbb{C}^m \backslash \mathbb{R}^m,$ the argument of the $\log$ is less than 1, and $\lambda_z$ is positive.  The formula for $E_{\gamma_{cx}}(Z_{h_N(z)} )$ is a special case of a result in \cite{ek}, and is a very simple function:
$$ E_{\gamma_{cx}}(Z_{h_N(z)} ) = \frac{mN^m}{\pi^m}\frac{1}{(1+||z||^2)^{m+1}}.$$
The formula for $E_{\gamma_{real}}(Z_{h_N(z)} ) $ is very complicated, but, by Theorem \ref{thm1}, we know that $E_{\gamma_{real}}(Z_{h_N(z)} ) $ equals a very simple function, $E_{\gamma_{cx}}(Z_{h_N(z)} )$, plus some exponentially small term.

 We prove this result using the Poincar\'{e}-Lelong formula, which is similar to that which was used in \cite{bszpl}, but has the added complication that the coefficients are real.  The proof uses 2-point Szeg\H{o} kernel asymptotics, which still applies to the polynomials with real coefficients because we are viewing them as functions of complex variables.  We also use the fact that the $f_{q,N}(z)$'s are independent, which is a major difference from the critical points case considered in \cite{bmcrit}.

 Shiffman and Zelditch \cite{SZdist} and Bleher, Shiffman, and Zelditch (\cite{bszpl}, and \cite{bszuniv}) have generalized many results about random polynomials on $\mathbb{C}^m$ and $\mathbb{R}^m$ to complex manifolds, and they have several results related to the statistics of zeros of a random holomorphic section of a power of a line bundle over a complex manifold.  In particular, in \cite{bszpl}, the authors use the Poincar\'{e}-Lelong formula to find a formula for the density of zeros and correlations between zeros.  Edelman and Kostlan used a similar approach in \cite{ek} to get a result like $E_{\gamma_{cx}}(Z_{h_N(z)} )$ but for systems of more general complex functions.

In \cite{dszcrit}, Douglas, Shiffman, and Zelditch look at the critical points of a holomorphic section of a line bundle over a complex manifold, motivated by applications in string theory.  They use a generalized Kac-Rice formula to find statistics of these critical points, namely the density of critical points and correlations between critical points.  In \cite{bmcrit}, we study complex critical points of a random polynomial with \textit{real} coefficients and generalize the result in Theorem \ref{thm1} to the density of critical points of a SO($m+1$) polynomial.

\subsection{Behavior of the scaled zero density near $\mathbb{R}^m$}  Consider $h _{m,N}(z)$ as above.  Note that the behavior of the density function in the $m=1$ case (Figure \ref{zeros1}) and the $m=2$ case (Figure \ref{zerosm}) differ greatly. Consider the scaling limit of the density, $$ K_{\gamma_{real}}^{\infty}(z)  = \lim_{N\rightarrow\infty}\frac{1}{N} E_{\gamma_{real}}(Z_{h_N(\frac{z}{\sqrt{N}})} ),$$ which will help us understand the behavior of the density function in a region around $\mathbb{R}^m$ that is shrinking at a rate of $\frac{1}{\sqrt{N}}.$ We can show that $K_{\gamma_{real}}^{\infty}(z)$ depends only on $ y = \im z ,$ so we can write the scaled density as $K_{\gamma_{real}}^{\infty}(y).$

Figure \ref{scaledzeros} illustrates the behavior of $K_{\gamma_{real}}^{\infty}(y)$ near $\mathbb{R}^m$ for $m=1$ and $m=2.$ Note that for $m=2,\, K_{\gamma_{real}}^{\infty}(y)$ does not tend linearly towards zero as in the $m=1$ case, but instead it tends to infinity.  We prove the following:
\begin{mainthm}\label{thm2}\addcontentsline{toc}{subsection}{Theorem \ref{thm2}}
For $y$ near 0, \begin{align*}
K_{\gamma_{real}}^{\infty}(y) &= O(|y|), \quad \quad \,\,\,\,\, m=1, \\
K_{\gamma_{real}}^{\infty}(y) &= O\left(\frac{1}{||y||^{m}}\right),\,\, m\geq2.
\end{align*}
\end{mainthm} \n We will see that difference in the $m=1$ and $m\geq2$ cases boils down to the fact that
\begin{align*} \frac{\partial^2}{\partial y^2} |y| &=0, \quad \quad \quad \quad \quad \text{ for }y \in\mathbb{R}\backslash \{0\}, \\ \frac{\partial^2}{\partial y_jy_k} ||y|| &= O\left(\frac{1}{||y||}\right), \quad \text{ for } y \in\mathbb{R}^m\backslash \{0\}, m\geq2.
\end{align*}

Finally, after working mostly on $\C\backslash\R$ and $\C^m\backslash\R^m,$ we give a weak limit for $E_{\gamma_{real}}(Z_{f_N}(z)) $ in the $m=1$ case.
 We show that $\frac{1}{N} E_{\gamma_{real}}(Z_{f_N}(z)) = \frac{1}{N} E_{\gamma_{cx}}(Z_{f_N}(z))  +O(N^{-1}), $ weakly on $K \subset \mathbb{C}.$
We stress that $K$ could contain some points in $\mathbb{R},$ whereas our strong convergence result excludes points in $\mathbb{R}.$

\setcounter{maintheoremnumber}{0}
%%%%%%%%%%%%%%%%%%%%%%%%%%%%%%%%%%%%%%%%%%%%%%%%%%%%%%%%%%%%%%%%%%%%%%%%%%%%%%%%%%
%\input{P-L-formula-1-variable-zeros}

\section{Proof of Theorem \ref{thm1} for $m=1$}

\addcontentsline{toc}{subsection}{An introduction and list of results used in the proof}
Consider the real random polynomial $$\dis f_N(z) = \sum_{\ell=0}^{N} \tilde{a}_\ell z^\ell,$$ where the $\tilde{a}_\ell$'s are
\textit{real} independent Gaussian random variable with mean 0 and variance $N\choose \ell$.  Alternatively, one often writes
$$\dis f_N(z) = \sum_{\ell=0}^{N} a_\ell {N\choose \ell} ^{1/2} z^\ell,$$ where $a_\ell$ is a standard real Gaussian random variable.  Instead, we choose to think of the random polynomial
$$ f_N(z) = \sum_{\ell=0}^{N} c_\ell {N\choose \ell} ^{1/2} z^\ell,$$ where $c_\ell$ is a more general complex random variable with associated measure $d\gamma$.  We then consider two special cases
\begin{align*}
d\gamma_{cx} &=
\frac{1}{\pi^{N}}e^{-|c|^2} dc, c \in \mathbb{C}^{N+1}, %D_N = {N+m \choose m}
\\
d\gamma_{real} &= \delta_S \frac{1}{\pi^{N}}e^{-|c|^2} dc, c \in \mathbb{C}^{N+1}, %D_N = {N+m \choose m}
\end{align*}  where $\delta_S$ is the delta function on $S \subset \mathbb{C}^{ N+1},$ the set of points $c=a+ib \in \mathbb{C}^{ N+1}$ where $b=0\in \mathbb{R}^{ N+1}.$  Here $d\gamma_{cx}$ corresponds to the standard complex Gaussian coefficients case, where we are considering
$$f_N(z) = \sum_{\ell=0}^{N} c_\ell {N\choose \ell} ^{1/2} z^\ell, $$
where the $c_\ell$'s are standard complex Gaussian random variables, and $d\gamma_{real}$ corresponds to the standard real Gaussian coefficients case, where we have
$$ f^{}_{N}(z) = \sum_{\ell=0}^{N} c_\ell {N\choose \ell} ^{1/2} z^\ell = \sum_{\ell=0}^{N} a_\ell {N\choose \ell} ^{1/2} z^\ell,$$
where $c_\ell = a_\ell +i0$ is a standard real Gaussian random variable.
We let $E_{\gamma}(\cdot)$ denote expectation with respect to $d\gamma_{}$.

 The goal of this section is to prove Theorem \ref{thm1} in the $m=1$ case:
\begin{proposition}[Theorem \ref{thm1} for $m=1$]\label{thm1_1}
We can write
\begin{align*}
E_{\gamma_{real}}(Z_{f_N}(z)) &=  E_{\gamma_{cx}}(Z_{f^{}_N}(z)) + \tilde{E}_N(z),
\end{align*}
where
\begin{align*}
E_{\gamma_{cx}}(Z_{f^{}_N}(z)) &= \frac{N}{\pi}\frac{1}{(1+|z|^2)^2}
\end{align*}
and where $\tilde{E}_N(z)=O(e^{-\lambda_z N}),$
for all $z \in \mathbb{C} \backslash \mathbb{R}.$  Here $ \lambda_z = -\log \big|\frac{1+z^2}{1+|z|^2} \big| $ is a positive
constant that depends continuously on $z$.
\end{proposition}
Using the Poincar\'e-Lelong formula, we write $E_{\gamma_{real}}(Z_{f_N}(z)) =  E_{\gamma_{cx}}(Z_{f^{}_N}(z)) + \tilde{E}_N(z)$, and we then aim to find an explicit formula for this error term.  Writing $f_N(z)= a \cdot F_N(z)$ and $u_N(z) = \frac{F_N(z)}{||F_N(z)||}$, we can write
    $$ \tilde{E}_N(z) \,\,dx \wedge dy = E_{\gamma_{real}}\left(\frac{i}{\pi} \partial \bar{\partial} \log |a \cdot u_N(z)|\right). $$
Note that $E_{\gamma_{real}}(\cdot)$ denotes an integral over all of $\mathbb{R}^N$ with respect to $d\gamma_{real}$ and is fairly complicated integral.  However, also note that $d\gamma_{real}$, the real standard Gaussian measure, is rotationally invariant.  We use this fact and perform 2 real orthogonal changes of variables in order to simplify this integral over $\mathbb{R}^N$ and write it as an integral over $\R^2$:
\begin{lemma}[Performing real rotations]\label{rotation1}
$$\tilde{E}_N(z)\,\,dx \wedge dy = \frac{i}{\pi}\partial \bar{\partial} \int_{\mathbb{R}^2}             \log | a_0 (r+is)+ a_1(it)| \frac{1}{2\pi} e^{-(a_0^2+a_1^2)/2} \,da_0 da_1$$
for some functions $r=r_N(z),\,s=s_N(z),$ and $t=t_N(z)$ for which we give formulas within the proof of the lemma.
\end{lemma}
We make another change of variables, this time switching $(a_0,a_1)$ to polar coordinates $(\rho,\theta)$.  We can easily integrate with respect to $\rho$, further simplifying our error term to an integral with respect to $\theta$ over $[0,2\pi].$ We apply Jensen's formula to what is left to evaluate the integral and get an explicit formula for $\tilde{E}_N(z)$:

\begin{lemma}[Evaluating the integral] \label{eval1} If $r^2+s^2+t^2 = 1$ and $r,s,t>0$, we have
$$ \frac{1}{2\pi} \int_{\mathbb{R}^2}             \log | a_0 (r+is)+ a_1(it)|  e^{-(a_0^2+a_1^2)/2} \,da_0 da_1 = \frac{1}{ 2 }  \log (1+2rt) .$$
\end{lemma}

\begin{lemma}[An exact formula for, and asymptotics for, the error term]\label{exact1}
We show that, by Lemma \ref{rotation1} and Lemma \ref{eval1}, and after some simplification of $2rt$, we have
\begin{align*}
\tilde{E}_N(z) =   \frac{1}{\pi} \frac{\partial^2}{ \partial z \partial \bar{z}} \log
\left(1+\sqrt{1-\left|\frac{(1+z^2)^N}{(1+|z|^2)^N}\right|^2}\right) , \,z \in \mathbb{C}\backslash\mathbb{R},
\end{align*}
It follows that $\tilde{E}_N(z) = O(e^{-\lambda_z N}), z \in \mathbb{C} \backslash \mathbb{R},$ where $\lambda_z= -\log \big|\frac{1+z^2}{1+|z|^2} \big| $ is a positive constant depending continuously on $z$.
\end{lemma}
\n This result gives us Theorem \ref{thm1} for the one variable case.

In subsections 1 through 5, we prove the aforementioned results.  The approach we use is similar to that described in \cite{bszpl}, where they find the limit of the pair correlations of zeros of
random holomorphic sections of powers of a line bundle of a complex manifold. While we only deal with density of zeros in this
section, the condition that the coefficients $a_j$ are real causes the method in \cite{bszpl} to be useful.

\subsection{Proof of Proposition \ref{thm1_1} - Theorem \ref{thm1} for $m=1$}
We write $a = (a_0, ..., a_N)$ and
$$F_N = \left({N\choose 0} ^{1/2} z^0,  {N\choose 1} ^{1/2} z^1, ...,  {N\choose N} ^{1/2} z^N \right),$$ so that $f_N= a \cdot F_N.$ By the
Poincare-Lelong formula, the density of the zeros of $f, E_{\gamma_{real}}(Z_{f_N}),$ satisfies
\begin{align}\label{plresult}
    E_{\gamma_{real}}(Z_{f_N})\,\,dx \wedge dy = E_{\gamma_{real}}(\frac{i}{\pi} \partial \bar{\partial} \log |f_N|) = E_{\gamma_{real}}(\frac{i}{\pi} \partial \bar{\partial} \log |a \cdot F_N|).
\end{align}
\n We write $F_N(z) = ||F_N(z)||u_N(z), $ where $u_N(z)$ is a unit vector, and (\ref{plresult}) becomes
\begin{align*}
    E_{\gamma_{real}}(\frac{i}{\pi} \partial \bar{\partial} \log ||F_N(z)||) + E_{\gamma_{real}}(\frac{i}{\pi} \partial \bar{\partial} \log |a \cdot u_N(z)|)
\end{align*}
From ~\cite{bszpl} we can see that $$E_{\gamma_{cx}}(Z_{f_N}(z)) \,\,dx \wedge dy= \frac{i}{\pi} \partial \bar{\partial} \log ||F_N(z)|| = E_{\gamma_{real}}(\frac{i}{\pi} \partial \bar{\partial} \log ||F_N(z)||)\,\,dx \wedge dy,$$ so we have
    $$E_{\gamma_{real}}(Z_f) = E_{\gamma_{cx}}(Z_f) + \tilde{E}_N(z). $$

\n We now prove the Lemmas, which deal with the error term $\tilde{E}_N(z)$, and return to the proof of Proposition \ref{thm1_1} in Section \ref{thm1_1b}.

\subsection{Proof of Lemma \ref{rotation1} - Real Rotations} As shown in \cite{bszpl}, when $a_j$ is a standard complex Gaussian random variable, this error term $\tilde{E}_N(z)$ is zero
\emph{for all} $N$ (not just as $N\rightarrow \infty).$  Because of the SU(2)-invariance of the standard complex Gaussian measure,
one can perform a unitary change of variables so that $u$ becomes $(1,0,...,0)$ and the integral $\int_{\mathbb{C}^{N+1}}
\partial \bar{\partial} \log |a \cdot u| d\gamma_{cx}(a)$ becomes a single integral that evaluates to 0:
$$\int_{\mathbb{C}^{N+1}}
\partial \bar{\partial} \log |a \cdot (1,0,...,0)| d\gamma_{cx}(a) = \int_{\mathbb{C}} \partial \bar{\partial}  \log |a_0 | d\gamma_{cx}(a_0) = 0.$$

In the case where $a_j$ is real, the second term is not zero for all $N.$ Because only real rotations can be performed, $u$ can
not be rotated to $(1,0,...,0),$ giving a single integral.  But we can still use the rotational invariance of real Gaussian
measures to obtain a double integral over $\mathbb{R}^2$, which is a little more manageable than the integral over
$\mathbb{R}^{N+1}.$

Let $u=\re u + i\im u= (\re u_1, ..., \re u_N) + i(\im u_1, ..., \im u_N).$  Note that $u, \re u,$ and $\im u$ depend on $z$ and
$N$ but we frequently omit these arguments for convenience. Since we need to do real rotations, the real and imaginary parts of
$u$ must be rotated the same.  Therefore, as mentioned, we can not rotate $u$ to (1, 0, ..., 0). However,
we can rotate so that either the real part or the imaginary part of $u$ is of the form $(r, 0, ..., 0),$ where $r=r_N(z)$ is some
(non-zero) constant less than 1. So we choose to perform a (real) rotation of $a_0, a_1, ...., a_N$ so that
$$\tilde{u} = \re \tilde{u} +i\im \tilde{u} = (r, 0,..., 0) + i(\im \tilde{u_1}, ..., \im \tilde{u_N}).$$
\n Then one can perform a rotation of the $a_1, ..., a_N$ variables so that $\re u$ is unaffected and $u$ becomes
\begin{align*}
&(r_N(z), 0,..., 0) + i(s_N(z), t_N(z), 0,...,0) \\
= &(r_N(z)+i s_N(z), i t_N(z), 0,...,0). \end{align*}
\n Note that since $u$ is a unit vector, and rotations preserve length, $r, s, $ and $t$ have the condition $r^2+s^2+t^2=1.$ Note
also that $r,s,$ and $t$ all depend on $z$ and $N$ but we frequently omit these.  We are now concerned with the limit of the
simpler integral,
\begin{align*}
\tilde{E}_N(z) = &\frac{i}{\pi}\int_{\mathbb{R}^{N+1}} \partial \bar{\partial}             \log |(a_0, a_1,...,a_N) \cdot (r+i s, i t, 0,...,0)|
\,d\gamma_{real}(a)
\\= &\frac{i}{\pi}\int_{\mathbb{R}^2} \partial \bar{\partial}             \log | a_0 (r+is)+ a_1(it)| \,d\mu(a_0) d\mu(a_1)
\\= &\frac{i}{\pi}\partial \bar{\partial} \int_{\mathbb{R}^2}             \log | a_0 (r+is)+ a_1(it)| \frac{1}{2\pi} e^{-(a_0^2+a_1^2)/2} \,da_0 da_1.
\end{align*}

\subsubsection{Formula for $r.$}
First, we know that since $u(z)=\frac{F(z)}{||F(z)||},$ and since the length of $\re u$ doesn't change from a rotation, we can
write
$$r^2=||\re \tilde{u}||^2 = ||\re u||^2 = \frac{||\re F||^2}{||F||^2} = \frac{1}{2} \,+\, \frac{1}{2}\,\re\left(\frac{1+z^2 }{1+|z|^2}\right)^N = \frac{1}{2} \,+ O(e^{-\lambda_z N}),$$ where $e^{-\lambda_z}=\big|\frac{1+z^2}{1+|z|^2}\big| <1, z \in \C\backslash\R $,  Note that we are assuming $\im z \neq 0,$ and note that $\lambda_z = - \log\Big|\frac{1+z^2}{1+|z|^2}\Big|>1.$

\subsubsection{Formula for $s.$} Next, we have the relationship $\re \tilde{u} \cdot \im \tilde{u} = rs,$ so since the angle between
$\re u$ and $\im u$ doesn't change under a rotation, we have
\begin{align*}
s = \frac{\re \tilde{u} \cdot \im \tilde{u}}{r} &=  \frac{\re u \cdot \im u}{r} = \frac{\re \frac {F}{||F||} \cdot \im \frac{F}{||F||}}{r} =
\frac{\re F \cdot \im F}{||F||^2} \cdot \frac{1}{r} \\ &= \frac{1}{2} \im \left(\frac{1+z^2}{1+|z|^2}\right)^N \cdot \frac{1}{r} = O(e^{-\lambda_z N}), z \in \C \backslash \R.
\end{align*}

\subsubsection{Formula for $t.$} Since $r^2+s^2+t^2=1,$ we have $t$ easily:  $$[t_N(z)]^2 = 1-[r_N(z)]^2-[s_N(z)]^2 = \frac{1}{2} + O(e^{-\lambda_z N}), z \in \C \backslash \R.$$

Also, we note that $r,s$, and $t$ converge uniformly on compact sets $K \subset \mathbb{C} \backslash
\mathbb{R},$ where the error term is $O(e^{-\lambda_K  N}),$ for some universal constant $\lambda$ (independent of $N$ and $z$).  Finally, we note that all derivatives of $r,s$, and $t$ are $O(e^{-\lambda_z N}),$ and all derivatives converge uniformly on compact sets $K \subset \mathbb{C} \backslash \mathbb{R}$ as well.

\subsection{Proof of Lemma \ref{eval1} - Evaluating the integral}
 We now prepare to use Jensen's formula to evaluate the integral
\begin{align*}
&\frac{i}{\pi}\partial \bar{\partial} \int_{\mathbb{R}^2}             \log | a_0 (r+is)+ a_1(it)| \frac{1}{2\pi}
e^{-(a_0^2+a_1^2)/2} \,da_0 da_1.
\end{align*}
\subsubsection{Switch to polar coordinates}  First, we switch to polar coordinates, so the integral becomes
\begin{align*}
& \frac{i}{2\pi^2} \partial \bar{\partial} \int_{\theta=0}^{2\pi} \int_{\rho=0}^{\infty}             \log |(\rho \sin \theta
)(r+is)+ (\rho\cos\theta)(it)| e^{-\rho^2/2}\,\rho d\rho d\theta.
\end{align*}
\n We may factor out a $\rho$ from the argument of the $\log$ and get
\begin{align*}
& \frac{i}{2\pi^2} \partial \bar{\partial} \int_{\theta=0}^{2\pi} \int_{\rho=0}^{\infty}     \left( \log \rho  +     \log |(\sin
\theta )(r+is)+ (\cos\theta)(it)| \right) e^{-\rho^2/2}\,\rho d\rho d\theta.
\end{align*}

\n Since $$\dis \int_{\theta=0}^{2\pi} \int_{\rho=0}^{\infty}   \log \rho e^{-\rho^2} \rho d\rho d\theta$$ doesn't depend on $z,$ it
gets killed by $\partial \bar{\partial},$ so we are left with
\begin{align*}
& \frac{i}{2\pi^2} \partial \bar{\partial} \int_{\theta=0}^{2\pi} \int_{\rho=0}^{\infty}     \log |(\sin \theta )(r+is)+
(\cos\theta)(it)| e^{-\rho^2/2}\,\rho d\rho d\theta.
\end{align*}

\n The $\log$ term doesn't depend on $\rho$, so we may pull that term outside the integral, and integrate with respect to $\rho$
to get
\begin{align*}
  & \frac{i}{2\pi^2} \partial \bar{\partial} \int_{\theta=0}^{2\pi}  \log |(\sin \theta )(r+is)+ (\cos\theta)(it)|  \left[\int_{\rho=0}^{\infty}   e^{-\rho^2/2}\,\rho \,d\rho \right] d\theta \\
= \,\,\,\,\,\,& \frac{i}{2\pi^2} \partial \bar{\partial} \int_{\theta=0}^{2\pi}  \log |(\sin \theta )(r+is)+ (\cos\theta)(it)|  \left[-  e^{-\rho^2/2} \right]_0^\infty \, d\theta \\
= \,\,\,\,\,\,& \frac{i}{2\pi^2} \partial \bar{\partial} \int_{\theta=0}^{2\pi}  \log |(\sin \theta )(r+is)+ (\cos\theta)(it)|  \left[1\right] \, d\theta \\
=& \frac{i}{2\pi^2} \partial \bar{\partial} \int_{\theta=0}^{2\pi}  \log |(\sin \theta )(r+is)+ (\cos\theta)(it)|  \, d\theta.
\end{align*}

\subsubsection{Jensen's Formula}
\n Using the fact that $\cos \theta = \frac{1}{2}( e^{i\theta} + e^{-i\theta})$ and $\sin \theta = \frac{1}{2i}( e^{i\theta} -
e^{-i\theta}).$, we can write
\begin{align*}
 -& \frac{i}{2\pi^2} \partial \bar{\partial} \int_{\theta=0}^{2\pi}  \log \frac{1}{2}|( e^{i\theta} - e^{-i\theta})(-i)(r+is)+ ( e^{i\theta} + e^{-i\theta})it|  \, d\theta.
\end{align*}
\n We can bring out a $\log \frac{1}{2}$, and since $\partial \bar{\partial} \log \frac{1}{2} = 0,$ we have

\begin{align*}
 -& \frac{i}{2\pi^2} \partial \bar{\partial} \int_{\theta=0}^{2\pi}  \log |( e^{i\theta} - e^{-i\theta})(-ir+s)+ ( e^{i\theta} + e^{-i\theta})it|  \, d\theta.
\end{align*}

\n We can factor out $e^{-i\theta},$ and since $\log|e^{-i\theta}|=0,$ we get
\begin{align*}
 -& \frac{i}{2\pi^2} \partial \bar{\partial} \int_{\theta=0}^{2\pi}  \log |( e^{i2\theta} - 1)(-ir+s)+ ( e^{i2\theta} + 1)it|  \, d\theta \\
 =-& \frac{i}{2\pi^2} \partial \bar{\partial} \int_{\theta=0}^{2\pi}  \log | (s+i(t-r)  )e^{i2\theta}+ (-s+i(t+r))|  \, d\theta.
\end{align*}
We can now use Jensen's formula to evaluate the inner integral.  Recall that Jensen's formula states that, assuming $\phi(0)
\neq0,$ and $\phi$ is non-zero on $\partial D(0,1),$ then
\begin{align*}
\frac{1}{2\pi} \int_{\theta=0}^{2\pi} \log |\phi(e^{i\theta})| d\theta = \log |\phi(0)| + \sum_{\phi(w_j)=0 \atop |w_j| < 1} \log
\frac{1}{|w_j|}.
\end{align*}

In our case, $$\phi(w) =  (s+i(t-r)  )w^2+ (-s+i(t+r)),$$ so that we have
\begin{align*}
\phi(e^{i\theta}) &=  (s+i(t-r))e^{i2\theta}+ (-s+i(t+r)) \\
\phi(0) &= -s+i(t+r), \text{ and } \\
\phi(w_j) &= 0 \iff [w_j(z)]^2 = -\frac{-s+i(t+r)}{s+i(t-r)}.
\end{align*}
Note that since $|\phi(0)|^2 = |-s+i(t+r)|^2 = s^2+(t+r)^2 = s^2+r^2+2rt+t^2  = 1+2rt$, and $r$ and $t$ are non-negative (by
construction), $\phi(0) \neq 0.$

We show that $|w_j(z)|^2 \geq 1,$ for all $z,$ implying that $|w_j(z)| \geq 1$ and that all the zeros $w_j(z)$ of $\phi$ are
outside the unit disk for every $z$.  We have
\begin{align*}
|w_j(z)|^4 = \Big|\frac{-s+i(t+r)}{s+i(t-r)}\Big|^2 = \frac{s^2+t^2+2rt+r^2}{s^2+t^2-2rt+r^2} \geq 1
\end{align*}
for $z \in \mathbb{C}$ since $r$ and $t$ are non-negative by construction.  Note that the only time $rt$ is zero is when $z \in
\mathbb{R}.$  In this case, $r_N(z) = 1$ and $s_N(z)=t_N(z)=0$ for all $N,$ and $|w_j(z)|=1.$

So since all of the zeros of $\phi$ are outside the unit disk, we have  for $z \in \mathbb{C} ,$
\begin{align*}
\frac{1}{2\pi} \int_{\theta=0}^{2\pi} \log |\phi(e^{i\theta})| d\theta   &= \log |\phi(0)|
                                         = \log |-s+i(t+r)| \\
                                        &= \frac{1}{2}\log |-s+i(t+r)|^2
                                         = \frac{1}{2}\log (1+2rt),
\end{align*}
or
\begin{align*}
\frac{1}{2\pi} \int \log |a_0(r+is)+a_1(it)| e^{-a_0^2 - a_1^2} da_0 da_1
                                         = \frac{1}{2}\log (1+2rt),
\end{align*}
the result of the lemma.
\subsection{Proof of Lemma \ref{exact1} - An exact formula for the error term}
From the proof of Lemma \ref{eval1}, we have
\begin{align*}
\tilde{E}_N(z)\,\,dx \wedge dy &= -\frac{i}{2\pi^2}     \partial \bar{\partial} \int \log |\phi(e^{i\theta})| d\theta
 =  -\frac{i}{2\pi^2}  \partial \bar{\partial}  \left(2\pi\right) \frac{1}{2}   \log (1+2rt) \\
&=  -\frac{i}{2\pi }  \partial \bar{\partial}  \log (1+2rt)
 =   \frac{1}{ \pi }  \frac{\partial^2}{ \partial z \partial \bar{z}}  \log (1+2rt) \,\,dx \wedge dy,
z\in\mathbb{C}\backslash\mathbb{R}.
\end{align*}
After some simplification of $2rt$ we have
\begin{align*}
\tilde{E}_N(z) = \frac{1}{ \pi }  \frac{\partial^2}{ \partial z \partial \bar{z}} \log
\left(1+\sqrt{1-\left|\frac{(1+z^2)^N}{(1+|z|^2)^N}\right|^2}\right)  , z\in\mathbb{C}\backslash\mathbb{R}.
\end{align*}

\subsection{Finishing Proof of Proposition \ref{thm1_1} - Theorem 1 for $m=1$ }\label{thm1_1b} Since we found that $r =r_N(z)= \sqrt{\frac{1}{2}} + O(e^{-\lambda_z N}),t =t_N(z)= \sqrt{\frac{1}{2}} + O(e^{-\lambda_z N}),$ and $s= s_N(z)= O(e^{-\lambda_z N})$ and all derivatives (in
particular, the first and second derivatives) of $r,s,$ and $t$ are $O(e^{-\lambda_z N})$, we can say that, from the proof of Lemma \ref{exact1},
\begin{align*}
\tilde{E}_N(z)
%&= E(\frac{i}{\pi}\partial \bar{\partial} \log |a \cdot u_N(z)|) \\&= -\frac{i}{2\pi }  \partial \bar{\partial} \log (1+2rt)
= O(e^{-\lambda_z N}), z\in \mathbb{C}
\backslash \mathbb{R}.
\end{align*} The proof follows directly from this result, combined with Lemma \ref{pl1} and Lemma \ref{exact1}.

%%%%%%%%%%%%%%%%%%%%%%%%%%%%%%%%%%%%%%%%%%%%%%%%%%%%%%%%%%%%%%%%%%%%%%%%%%%%%%%%%%
%\input{P-L-formula-m-variables-zeros}

\section{Proof of Theorem \ref{thm1} for $m \geq 2$}
\addcontentsline{toc}{section}{An introduction and list of results used in the proof}
In this section we are concerned with the zeros of $h _{m,N}=(f_{1,N}, ..., f _{m,N}):\mathbb{C}^m \rightarrow \mathbb{C}^m $, where
$f _{q,N}$ is a polynomial of the form
\begin{align*}
f _{q,N}(z) &= \sum_{|J|=0}^{N} a_{J}^q {N\choose J} ^{1/2} z^J
\end{align*}
where $a_{J}^q$ is a \textit{real} standard Gaussian random variable, and where we use the following multi-index notation:
\begin{align*}
z &= (z_1,...,z_m)\\
|J| &= j_1+ \cdots + j_m \\
a^q_J &= a^q_{j_1...j_m} \in \mathbb{R}\\
{N\choose J} &=  {N\choose j_1, ..., j_m} = \frac{N!}{(N-j_1-...-j_m)!j_1!\,\,...\,\,j_m!} \\
z^J &= z_1^{j_1}...z_m^{j_m}. \\
\end{align*}

%%%%%%
Instead, we choose to think of the random polynomials
$$ f_{q,N}(z) = \sum_{|J|=0}^{N} c_J^q {N\choose J} ^{1/2} z^J,$$ where the $c^q_J$'s are independent complex random variables, where the random vector $\{c^q_J\}\in D_N, D_N = {N+m \choose m}$ has associated measure $d\gamma$ for each $q$.  We then consider two special cases
\begin{align*}
d\gamma_{cx} &=
\frac{1}{\pi^{N}}e^{-|c|^2} dc, c \in \mathbb{C}^{D_N}, D_N = {N+m \choose m}
\\
d\gamma_{real} &= \delta_S \frac{1}{\pi^{N}}e^{-|c|^2} dc, c \in \mathbb{C}^{D_N},
\end{align*}  where $\delta_S$ is the delta function on $S \subset \mathbb{C}^{D_N},$ the set of points $c=a+ib \in \mathbb{C}^{D_N}$ where $b=0\in \mathbb{R}^{D_N}.$  Here $d\gamma_{cx}$ corresponds to the standard complex Gaussian coefficients case, where we are considering
$$f_{q,N}(z) = \sum_{|J|=0}^{N} c^q_J {N\choose J} ^{1/2} z^J, $$
where the $c^q_J$'s are standard complex Gaussian random variables, and $d\gamma_{real}$ corresponds to the standard real Gaussian coefficients case, where we have
$$ f^{}_{q,N}(z) = \sum_{|J|=0}^{N} c^q_J {N\choose J} ^{1/2} z^J = \sum_{|J|=0}^{N} a^q_J {N\choose J} ^{1/2} z^J,$$
where $c^q_J = a^q_J +i0$ is a standard real Gaussian random variable.
We let $E(\cdot)$ denote expectation with respect to $d\gamma_{real}$ and $E_{\gamma_{cx}}(\cdot)$ denote expectation with respect to $d\gamma_{cx}$.
%%%%%%

The goal of this chapter is to prove Theorem \ref{thm1} in the $m\geq 1$ case.

\begin{mainthm}\setcounter{theorem}{0}
 \begin{align*} \dis E_{\gamma_{real}}(Z_{h_{m,N}(z)} ) &=  E_{\gamma_{cx}}(Z_{h_{m,N}(z)} ) +O(e^{-\lambda_z N}),
\end{align*}
for all $z \in \mathbb{C}^m \backslash \mathbb{R}^m,$ where $\lambda_z$ is a positive
constant that depends continuously on $z$.  The explicit formula for $\lambda_z$ is
$$ \lambda_z = - \log\Big|\frac{1+z\cdot z}{1+||z||^2}\Big|. $$
Also, for compact sets $K \subset \mathbb{C}^m\backslash\mathbb{R}^m$, the density converges uniformly with an error term of $O(e^{-\lambda_K N})$, where $\lambda_K$ is a constant that depends only on $K$.
\end{mainthm}
\n So at any point away from $\mathbb{R}^m$, the expected density of zeros in the real coefficients case approaches the
expected density of zeros in the complex coefficients case as $N$ gets large.

Using the Poincar\'e-Lelong formula, we can write
\begin{align*}
E_{\gamma_{real}}(Z_{h^{}_N}(z)) = E_{\gamma_{cx}}(Z_{h_N}(z)) + \tilde{E}_{N}(z)
\end{align*}
where
\begin{align*}
E_{\gamma_{cx}}(Z_{h_N}(z)) &= \frac{mN^m}{\pi^m}\frac{1}{(1+||z||^2)^{m+1}}  ,
\end{align*}
and where $\tilde{E}_{ N}(z)$
is some ``error term."  We then find an explicit formula for this error term.  Writing $f_{q,N}(z)= a \cdot F_{q,N}(z)$ and $u_{q,N}(z) = \frac{F_{q,N}(z)}{||F_{q,N}(z)||}$, we can write $ \tilde{E}_N(z) \,\,dx \wedge dy $ as a sum of $2^{m}-1$ terms, each of which contains a factor of the form
\begin{align}\label{keyfactor}
E_{\gamma_{real}}\left(\frac{1}{\pi} \frac{\partial^2}{\partial z_j \partial \bar{z_k}} \log |a \cdot u_N(z)|\right)
\end{align}
Note that $E_{\gamma_{real}}(\cdot)$ denotes an integral over all of $\mathbb{R}^{D_N},$ with respect to $d\gamma_{real}$, and is a fairly complicated integral.  However, like in the one variable case, we note that $d\gamma_{real}$ is rotationally invariant, and we use this fact and perform 2 real orthogonal changes of variables in order to simplify this integral over $\mathbb{R}^N$ and write it as an integral over $\R^2$:
\setcounter{lemma}{1}
\begin{lemma}[Lemma \ref{rotation1} for $m\geq 1$]\label{rotationm}  The factor (\ref{keyfactor}) equals
$$\frac{1}{\pi}  \int_{\mathbb{R}^2}    \log | a_0 (r+is)+ a_1(it)| \frac{1}{2\pi} e^{-(a_0^2+a_1^2)/2} \,da_0 da_1$$
for some functions $r=r_N(z),\,s=s_N(z),$ and $t=t_N(z)$ for which we give formulas within the proof of the lemma.
\end{lemma}
\n Then we easily get
\begin{lemma} [Lemma \ref{eval1} for $m\geq 1$]\label{evalm} . If $r^2+s^2+t^2 = 1$ and $r,s,t>0$, we have that
$$ \frac{1}{2\pi} \int_{\mathbb{R}^2}             \log | a_0 (r+is)+ a_1(it)|  e^{-(a_0^2+a_1^2)/2} \,da_0 da_1 = \frac{1}{ 2 }  \log (1+2rt) .$$
\end{lemma}
\n This means that the factors (\ref{keyfactor}) are of the form
$$\frac{1}{\pi} \frac{\partial^2}{\partial z_j \partial \bar{z_k}} \log (1+2rt)$$ each of which is $O(e^{-\lambda_zN})$ for some positive constant $\lambda_z$ depending continuously on $z$.  Using an argument that relies on the independence of the $f_{q,N}(z)'s$, we get the following:
\begin{lemma} [Lemma \ref{exact1} for $m\geq 1$]\label{exactm}
We give an exact formula for $\tilde{E}_{ N}(z),$ and show that it goes to zero rapidly, i.e.,
\begin{align*}
\tilde{E}_{ N}(z) =O(e^{-\lambda_zN}), \quad z \in \mathbb{C}^m \backslash \mathbb{R}^m,
\end{align*}
where $\lambda_z$ is a positive constant depending continuously on $z$.  We give an explicit formula for $\lambda_z$ in the proof of the theorem.
\end{lemma}
This lemma gives us our first main theorem, as shown in Section \ref{thm1proofb}.

\subsection{Proof of Theorem \ref{thm1}}\label{plm}
We begin by writing
\begin{align*}
a^q &= (a_{0,...,0}^q, ...,a_J^q, ...,a_{0,...,0,N}^q) \in \mathbb{R}^{D_N}\\
F_{q,N}(z) &= F_N(z) = \left({N\choose 0,...,0} ^{1/2}, ...,  {N\choose J} ^{1/2} z^J,...,  {N\choose 0,...,0,N} ^{1/2} z_1^0...z_m^N \right) \in \mathbb{C}^{D_N}\\
\end{align*}
where $D_N = {N+m \choose m},$ so that we can write $f _{q,N}= a^q \cdot F_{q ,N}= a^q \cdot F_N.$

By the Poincar\'e-Lelong formula, we have
\begin{align*}
E(Z_{f _1}(z) \wedge ... \wedge Z_{f _m}(z))
&= E(\frac{i}{2\pi} \partial \bar{\partial} \log |f_1|^2 \wedge ... \wedge \frac{i}{2\pi} \partial \bar{\partial} \log |f_m|^2) \\
&=E\Big[ \left(\frac{i}{2\pi}\right)^m \Big(\partial \bar{\partial} \log |a^1 \cdot F|^2 \wedge ... \wedge \partial
\bar{\partial} \log |a^m \cdot F|^2 \Big) \Big] %
\end{align*}
which we can write more succinctly as
\begin{align}\label{eq312}
\left(\frac{i}{2\pi}\right)^m E\Big(\bigwedge_{q=1}^m \partial \bar{\partial} \log \big|a^q \cdot F \big|^2 \Big) .%
\end{align}
Writing $F = \frac{F}{||F||} ||F||$ and $u = \frac{F}{||F||},$ we can write (\ref{eq312}) as
\begin{align}\label{eq313}
\notag &\left(\frac{i}{2\pi}\right)^m E\Big(\bigwedge_{q=1}^m \partial \bar{\partial} \log \big|a^q \cdot \frac{F}{||F||}||F|| \big|^2
\Big)\\%
  =&\left(\frac{i}{2\pi}\right)^m E\Big[\bigwedge_{q=1}^m \big(\partial \bar{\partial} \log ||F||^2 + \partial
\bar{\partial} \log |a^q \cdot u|^2 \big)\Big]%
\end{align}
Since $\partial \bar{\partial} \log ||F||^2 + \partial \bar{\partial} \log |a^q \cdot u|^2 $ is independent of $a^\ell$ for $l\neq
q,$ then we can write the expected value of the wedge product as the wedge product of the expected values, and (\ref{eq313}) becomes
\begin{align*}
 &\left(\frac{i}{2\pi}\right)^m \bigwedge_{q=1}^m E\Big[\partial \bar{\partial} \log ||F||^2 + \partial
\bar{\partial} \log |a^q \cdot u|^2 \Big]\\%
= &\left(\frac{i}{2\pi}\right)^m \bigwedge_{q=1}^m \left( \partial \bar{\partial} \log ||F||^2 + E\Big[\partial
\bar{\partial} \log |a^q \cdot u|^2 \Big]\right).\\%
\end{align*}
At this point, we could find the large $N$ limit;  we have essentially reduced the $m$-variables case to showing that
$E\Big[\partial \bar{\partial} \log |a^q \cdot u|^2 \Big]$  is $O(e^{-\lambda_z N}).$  If this term is indeed $O(e^{-\lambda_z N})$, then all but one term in the wedge product goes to zero exponentially fast.  Since we want an exact formula for the density of zeros, we delay the proof of the large $N$ limit, and we first work out the details of writing an exact formula more explicitly.  From that formula, the large $N$ limit and the scaling limit will follow easily.

We write
\begin{align*}
\left(\frac{i}{2\pi}\right)^m E\Big[\bigwedge_{q=1}^m \big(\partial \bar{\partial} \log ||F||^2 + \partial \bar{\partial} \log
|a^q \cdot u|^2 \big)\Big] \\ = [E_{1,N}(z) + E_{2,N}(z) + ... +E_{2^m,N}(z)] d\omega
\end{align*}

where \begin{align*} %
E_{1,N}(z) d\omega:=\left(\frac{i}{2\pi}\right)^m E\Big( \partial \bar{\partial} \log ||F||^2       &\wedge \partial \bar{\partial} \log ||F||^2 \wedge ... \wedge \partial \bar{\partial} \log ||F||^2    \Big) \\ %
E_{2,N}(z) d\omega:=\left(\frac{i}{2\pi}\right)^m E\Big( \partial \bar{\partial} \log |a^1 \cdot u| &\wedge \partial \bar{\partial} \log ||F||^2 \wedge ... \wedge \partial \bar{\partial} \log ||F||^2    \Big) \\ %
&\vdots \\
E_{2^m,N}(z) d\omega:=\left(\frac{i}{2\pi}\right)^m E\Big( \partial \bar{\partial} \log |a^1 \cdot u| &\wedge \partial \bar{\partial}\log |a^2 \cdot u|  \wedge ... \wedge \partial \bar{\partial} \log |a^m \cdot u|     \Big) \\ %
\end{align*}
\n We look at these $2^m$ terms and we claim that only the first term is non-zero in the limit.  The first term is known:
\begin{align*} %
E_{1,N}(z)  d\omega&= E_{\gamma_{cx}}(Z_{h_N}) = \left(\frac{i}{2\pi}\right)^m E\Big( \partial \bar{\partial} \log ||F||^2       \wedge \partial \bar{\partial} \log ||F||^2 \wedge ... \wedge \partial \bar{\partial} \log ||F||^2    \Big) %
\\&=   \frac{mN^m}{\pi^m}\frac{1}{(1+|z_1|^2+...+|z_m|^2)^{m+1}} d\omega
   =   \frac{mN^m}{\pi^m}\frac{1}{(1+||z||^2)^{m+1}} d\omega
\end{align*}
We let the error term $\tilde{E}_N(z)$ be the sum of the remaining terms:
$$\tilde{E}_N(z) = E_{2,N}(z) + ... +E_{2^m,N}(z).$$
We now prove the lemmas, and continue the proof of Theorem \ref{thm1} in Section \ref{thm1proofb}.
\subsection{Proof of Lemma \ref{rotationm}}
Consider the integral
$$\int _{\mathbb{R}^{D_N}}  \log |a^\ell\cdot u_N(z)| \,\,d\mu(a^\ell) .$$
Note that the terms $E_{2,N}(z), ... ,E_{2^m,N}(z)$ each contain a factor of this form.  Also note that this integral is in the same form as the error term in the one variable case (see Lemma \ref{rotation1}), so we proceed in a similar manner. We
rotate $ (a_{0\cdots 0}, ..., a_{0 \cdots 0N}) $ and then  $ (a_{10\cdots 0}, ..., a_{0\cdots 0N}) $ so that the integral becomes
\begin{align*}
&\int_{\mathbb{R}^{D_N}}  \log |a^\ell \cdot (r+is, it, 0,...,0)|  \,d\mu(a^\ell) \\& \quad \quad =  \int_{\mathbb{R}^2}  \log |a_{0\cdots 0} (r+is) + a_{10\cdots 0} it |  \,d\mu(a_{0\cdots 0})\,d\mu(a_{10\cdots 0})  %
\end{align*}
where $r=r_N(z),  s=s_N(z),$ and $t=t_N(z).$  We have similar formulae and large $N$ limits for $r,s,$ and $t$:
\begin{align*}
[r_N(z)]^2 &= \frac{1}{2} + \frac{1}{2} \re \left(\frac{1+z_1^2+\cdots+z_m^2}{1+|z_1|^2+\cdots+|z_m|^2}\right)^N = \frac{1}{2} +
O(e^{-\lambda_z N}), z \in \mathbb{C}^m \backslash \mathbb{R}^m ,
 \\%
s_N(z) &=               \frac{1}{2} \im \left(\frac{1+z_1^2+\cdots+z_m^2}{1+|z_1|^2+\cdots+|z_m|^2}\right)^N \frac{1}{r_N(z)}   =
O(e^{-\lambda_z N}), z \in \mathbb{C}^m \backslash \mathbb{R}^m ,
\\ %
[t_N(z)]^2 &= 1 - [r_N(z)]^2 - [s_N(z)]^2 =  \frac{1}{2} + O(e^{-\lambda_z N}), z \in \mathbb{C}^m \backslash \mathbb{R}^m  .
\end{align*}
Also, all derivatives of $r,s$ and $t$ are $O(e^{-\lambda_z N}), z \in \mathbb{C}^m \backslash \mathbb{R}^m.$  Finally, we note that,  $r,s,t$ and their derivatives converge uniformly on compact sets $K \subset \mathbb{C}^m \backslash
\mathbb{R}^m,$ where the error term is $O(e^{-\lambda_K  N})$ for some universal constant $\lambda$ (independent of $N$ and $z$).

\subsection{Proof of Lemma \ref{evalm}}

By Lemma \ref{eval1}, we have
\begin{align*}
\frac{1}{2\pi}\int_{\mathbb{R}^2}  \log |a_{0\cdots 0} (r+is) + a_{10\cdots 0} it |  \,d\mu(a_{0\cdots 0})\,d\mu(a_{10\cdots 0})  = \frac{1}{2}
\log (1 + 2rt).
\end{align*}

\subsection{Proof of Lemma \ref{exactm}}

Consider, $E_q$ the $q$-th term in $\tilde{E}_N(z)$. This term is of the form

\begin{align*} %
E_{q,N}(z) d\omega &= E(\partial \bar{\partial}  \phi_1^q \wedge ... \wedge \partial \bar{\partial} \phi_m^q) \\
\end{align*}
where $\phi_{l,N}^q(z)$ is either $\log ||F_N(z)||$ or $\log |a^\ell\cdot u_N(z)|$ for each $\ell.$  For example, for $E_{2,N}(z)$ we
have $ \phi_1^2 = \log |a^1 \cdot u|$ and $\phi_\ell^2 = \log ||F_N(z)||$ for $1<\ell\leq k$. Writing out the wedge product we get
\begin{align*} %
E_{q,N}(z) &= E\left[ \sum_{\sigma, \tau} (-1)^{\sigma+\tau}  \left(\frac{\partial^2}{\partial z_{\sigma(1)} \partial \bar{z}_{\tau(1)}}   \phi_1^q\right) \cdots %
                                   \left(\frac{\partial^2}{\partial z_{\sigma(m)} \partial \bar{z}_{\tau(m)}}   \phi_m^q\right) \,\, \right]\\ %
\end{align*}
where the sum is over all permutations $\sigma$ and $\tau$ of $\{1,2,...,m\},$ and where $(-1)^\sigma$ denotes the sign
associated to the permutation $\sigma.$ Since the sum is finite, we can write
\begin{align*} %
E_{q,N}(z) &= \sum_{\sigma, \tau} (-1)^{\sigma+\tau} E\left[  \left(\frac{\partial^2}{\partial z_{\sigma(1)} \partial \bar{z}_{\tau(1)}}   \phi_1^q\right) \cdots %
                                   \left(\frac{\partial^2}{\partial z_{\sigma(m)} \partial \bar{z}_{\tau(m)}}   \phi_m^q\right) \,\, \right] %
\end{align*}
or
\begin{align*} %
E_{q,N}(z) &= \sum_{\sigma, \tau} (-1)^{\sigma+\tau} E_{q,N}^{\sigma, \tau}(z)\\
\text{where   }\,\, E_{q,N}^{\sigma, \tau}(z)&:= E\left[  \prod_{\ell=1}^m \left(\frac{\partial^2}{\partial z_{\sigma(\ell)} \partial \bar{z}_{\tau(\ell)}}   \phi_\ell^q(z)\right)  \,\,   \right] %
\end{align*}
To simplify notation even more, let $$D_\ell^{\sigma,\tau} = \frac{\partial^2}{\partial z_{\sigma(\ell)} \partial \bar{z}_{\tau(\ell)}}
$$ so that we have
\begin{align*} %
E_{q,N}^{\sigma, \tau}(z) &= E\left[ \prod_{\ell=1}^m D_\ell^{\sigma,\tau}   \phi_\ell^q(z)  \,\,   \right] %
\end{align*}

Now, note that $\phi_\ell^q(z)$ does not depend on all of $a^1,...,a^m,$ but only depends \textit{at most} on $a^\ell.$  (If
$\phi_\ell^q(z) = \log ||F_N(z)||,$ then it doesn't depend on $a^\ell$ either.)  So because $\phi_\ell^q(z)$ is independent of $a^{\ell'}$
for all $\ell' \neq \ell,$ we will write this integral over $\mathbb{R}^{D_N} \times \cdots   \times \mathbb{R}^{D_N}$ as a product of
integrals over $\mathbb{R}^{D_N}.$ To do this, we first let
    $$L_q = \{\ell: \phi^q_\ell  \text{ is of the form } \log ||F_N(z)||\} \subset \{1, \ldots, m\}.$$ Note that $\phi^q_\ell  \text{ is
of the form } \log |a^\ell\cdot u_N(z)|\},$ for all $\ell\notin L_q$. We can now we split the product to get
\begin{align*} %
 E_{q,N}^{\sigma, \tau}(z)    &=  E \left[  \left(  \prod_{\ell \in L_q}  D_\ell^{\sigma,\tau}   \phi_\ell^q(z)   \right)   \left( \prod_{\ell\notin L_q}   D_\ell^{\sigma,\tau}   \phi_\ell^q(z) \right)    \right]\\%
                &= E \left[  \left( \prod_{\ell \in L_q}   D_\ell^{\sigma,\tau}   \log ||F_N(z)||    \right) \left(  \prod_{\ell\notin L_q}    D_\ell^{\sigma,\tau}   \log |a^\ell\cdot u_N(z)| \right)  \right]   %
\end{align*}
By the definition of expected value, we have
\begin{align*} %
E_{q,N}^{\sigma, \tau}(z) &=  \int _{\mathbb{R}^{D_N}} \left[  \left( \prod_{\ell \in L_q}   D_\ell^{\sigma,\tau}   \log ||F_N(z)||    \right) \left(  \prod_{\ell\notin L_q}    D_\ell^{\sigma,\tau}   \log |a^\ell\cdot u_N(z)| \right)  \right]  d\mu(a^1)\cdots d\mu(a^m), %
\end{align*}
Note that the first product is independent of $a^\ell$ for all $\ell\notin L_q,$ and the second product is independent of all $\ell \in L_q,$
so we can write $E_{q,N}^{\sigma, \tau}(z)$ as
\begin{align*} %
     \left[ \int _{\mathbb{R}^{|L_q|D_N}}   \prod_{\ell \in L_q}   D_\ell^{\sigma,\tau}  \log ||F_N(z)||  d\mu(a^\ell) \right] \left[ \int _{\mathbb{R}^{(m-|L_q|)D_N}}      \prod_{\ell\notin L_q}    D_\ell^{\sigma,\tau}   \log |a^\ell\cdot u_N(z)| \,\,d\mu(a^\ell)    \right]  %
\end{align*}
The first product is also independent of $a^\ell$ for all $\ell \in L_q,$ and since $\int_{\mathbb{R}^{D_N}} d\mu(a^\ell) = 1,$ we have for
the first integral
\begin{align*} %
 \prod_{\ell \in L_q}   D_\ell^{\sigma,\tau}  \log ||F_N(z)|| %
\end{align*}
Even more, the $\ell$-th factor in the second product depends only on $\ell$ and is therefore independent of all $\ell' \notin L$ not equal
to $\ell.$  So the integral of this product becomes a product of the integrals:
\begin{align*} %
          \prod_{\ell\notin L_q}    \int _{\mathbb{R}^{D_N}}  D_\ell^{\sigma,\tau}   \log |a^\ell\cdot u_N(z)| \,\,d\mu(a^\ell)     %
\end{align*}
and we can  switch the derivatives and the integral to get
\begin{align*} %
          \prod_{\ell\notin L_q}  D_\ell^{\sigma,\tau}   \int _{\mathbb{R}^{D_N}}    \log |a^\ell\cdot u_N(z)| \,\,d\mu(a^\ell) .    %
\end{align*}
Putting everything together we have
\begin{align*} %
 E_{q,N}^{\sigma, \tau}(z) =   \left[  \prod_{\ell \in L_q}   D_\ell^{\sigma,\tau}  \log ||F_N(z)||  \right]  \left[      \prod_{\ell\notin L_q}    D_\ell^{\sigma,\tau}   \int _{\mathbb{R}^{D_N}}  \log |a^\ell\cdot u_N(z)| \,\,d\mu(a^\ell)    \right]  %
\end{align*}

\n Lemma \ref{evalm} gives us
\begin{align*} %
 E_{q,N}^{\sigma,\tau}(z) =        \left[  \prod_{\ell \in L_q}   D_\ell^{\sigma,\tau}  \log ||F_N(z)||   \right] \left[      \prod_{\ell\notin L_q}    D_\ell^{\sigma,\tau}  \frac{1}{2} \log (1 + 2rt)    \right]  %
\end{align*}
\subsubsection{Exact formula} Further simplification gives
\begin{align*} %
 E_{q,N}^{\sigma,\tau}(z) =        \left[ \prod_{\ell \in L_q}   D_\ell^{\sigma,\tau}  \frac{1}{2} \log (1+||z||^2)^N   \right]  \left[      \prod_{\ell\notin L_q}    D_\ell^{\sigma,\tau}  \frac{1}{2} \log \left(1 + \sqrt{1- \left| \frac{ (1+z \cdot z)^N}{(1+||z||^2)^N }\right|^2}\right)    \right]  %
\end{align*}
where we use the notation $z \cdot z = z_1^2+ \cdots + z_m^2, ||z||^2 =|z_1|^2+ \cdots + |z_m|^2.$ Finally,
\begin{align*} %
E_{q,N} (z) = \sum_{\sigma, \tau} (-1)^{\sigma+\tau} E_{q,N}^{\sigma,\tau}(z) %
\end{align*}
and
\begin{align*} %
\tilde{E}_{N} (z) = \sum_{q=2}^{2^m} E_{q,N} (z).
\end{align*}

\subsubsection{Limit formula}

All derivatives of $\log ||F|| $ are bounded.  Next, $r =r_N(z)= \sqrt{\frac{1}{2}} + O(e^{-\lambda_zN}),$ $t =t_N(z)=
\sqrt{\frac{1}{2}} + O(e^{-\lambda_zN}),$ and $s= s_N(z)= O(e^{-\lambda_zN})$ and all derivatives (in particular, the first and
second derivatives) of $r,s,$ and $t$ are $O(e^{-\lambda_zN})$  on $ \mathbb{C}^m \backslash \mathbb{R}^m.$ So we can say that
all second derivatives of $\log (1+2r t )$ are $O(e^{-\lambda_zN})$ on   $ \mathbb{C}^m \backslash \mathbb{R}^m.$ This means that

\begin{align*}
E_{q, N}^{\sigma,\tau}(z ) =   O(e^{-\lambda_zN}), z \in \mathbb{C}^m \backslash \mathbb{R}^m.
\end{align*}
Since this is true for each $i,\sigma,$ and $\tau,$ we have
\begin{align*}
\tilde{E}_{N}(z ) = \sum_{q=2}^{2^m} \sum_{\sigma, \tau} (-1)^{\sigma+\tau} E_{q,N}^{\sigma,\tau}(z) &= O(e^{-\lambda_zN}), z \in
\mathbb{C}^m \backslash \mathbb{R}^m.
\end{align*}

\subsection{Finishing Proof of Theorem \ref{thm1}}\label{thm1proofb}
Using our work in Section \ref{plm} with Lemma \ref{exactm}, we get our main result:
\begin{align*}
E_{\gamma_{real}}(Z_{h_N(z)})= E_{\gamma_{cx}}(Z_{h_N(z)}) + \tilde{E}_{N}(z ) &= \frac{mN^m}{\pi^m}\frac{1}{(1+||z||^2)^{m+1}} + O(e^{-\lambda_z N}),z \in \mathbb{C}^m \backslash \mathbb{R}^m.
\end{align*}

%%%%%%%%%%%%%%%%%%%%%%%%%%%%%%%%%%%%%%%%%%%%%%%%%%%%%%%%%%%%%%%%%%%%%%%%%%%%%%%%%%
%\input{Scaled-zeros-1-variable-and-m-variables}
\section{Proof of Theorem \ref{thm2}}

\addcontentsline{toc}{subsection}{An introduction and list of results used in the proof}

We then wish to study the behavior of $E_{\gamma_{real}}(Z_{f_N}(z))$ near the real line.  We define the scaling limit of $E_{\gamma_{}}(Z_{f_N}(z))$ to be
$$K^\infty_{\gamma_{}}(z) =  \lim_{N\rightarrow\infty} \frac{1}{N} E_{\gamma_{}}(Z_{f(\frac{z}{\sqrt{N}})}) $$
The scaling limit helps us understand the behavior of the density function in a region around $\mathbb{R}$ that is shrinking at a rate of $\frac{1}{\sqrt{N}}.$  We note that $$K^\infty_{\gamma_{cx}}(z) = \frac{1}{\pi} ,$$ and we find the scaling limit of the error term when $m=1$:
\begin{lemma}[Scaling limit for the error term, $m=1$]\label{scalederror1}
\begin{align*}
\tilde{E}_\infty(z) = \lim_{N\rightarrow\infty} \frac{1}{N} \tilde{E}_N(\frac{z}{\sqrt{N}}) &= \frac{1}{  \pi } \frac{\partial^2}{ \partial z \partial
\bar{z}}  \log \left(1+\sqrt{1-\left|\frac{e^{z^2}}{e^{|z|^2}}\right|^2}\right)  \,\,,
z\in\mathbb{C}\backslash\mathbb{R}. %
\end{align*} By setting $z=x+iy,$ we can write
\begin{align*}
\lim_{N\rightarrow\infty} \frac{1}{N} \tilde{E}_N(\frac{z}{\sqrt{N}})&=  \frac{1}{ 4 \pi } \frac{\partial^2}{ \partial y^2 }  \log
\left(1+\sqrt{1-e^{-4y^2}}\right)  \,\,, y \neq 0.  %
\end{align*}
\end{lemma}
Then, by Proposition \ref{thm1_1} and Lemma \ref{scalederror1}, we recover Prosen's scaled density equation, and our Theorem \ref{thm2} for the special case $m=1$:
\begin{proposition}[Equation (26) in \cite{prosen}, and Theorem \ref{thm2} for $m=1$]\label{scaled1} We have
\begin{align*}
K^\infty_{\gamma_{real}}(z) = \frac{1}{\pi} \,
\frac{1-(4y^2+1)e^{-4y^2}}{(1-e^{-4y^2})^{3/2}}.
\end{align*}
Since $K^\infty_{\gamma_{real}}(z)$ depends only on $y$, we can write $K^\infty_{\gamma_{real}}(y)$, and we have the asymptotics
\begin{align*}
K^\infty_{\gamma_{real}}(y) = O(|y|)
\end{align*} for $y$ near zero.
\end{proposition}
\n Since $y=0$ corresponds to the real line, this result tells us that the scaled density tends linearly toward 0 as we approach the real line.  This formula for $K^\infty_{\gamma_{real}}(z)$ was given by Prosen as mentioned, but we find it here using Poincar\'{e}-Lelong method that will be generalized for the $m \geq 2$ case.  We give a formula for the scaling limit of the ``error term" $\tilde{E}_{ N}(z),$ when $z \in \C^m\backslash\R^m$, which we denote $\tilde{E}_{ \infty}(z), z \in \C^m\backslash\R^m$, and the scaling limit of the density, which we denote, $K_{\gamma_{real}}^{\infty}(y)$. (Once again, the scaled density only depends on $y= \im z$.)  We show that the behavior of $K_{\gamma_{real}}^{\infty}(y)$ near $\mathbb{R}^m$ for $m\geq 2$ is different than the behavior for $m=1$:
\addtocounter{theorem}{1}
\begin{mainthm}
For $y$ near 0, \begin{align*}
K_{\gamma_{real}}^{\infty}(y) &= O(|y|), \quad \quad \,\,m=1 \\
K_{\gamma_{real}}^{\infty}(y) &= O(\frac{1}{||y||^{m}}),\quad m\geq2.
\end{align*}
\end{mainthm}
We also show that as
$|\im z| \rightarrow \infty, \tilde{E}_{ \infty}(z) \rightarrow 0,$ so that $K_{\gamma_{real}}^{\infty}(y) \rightarrow 0, $ as $y\rightarrow \infty.$ In other words, the scaled density of zeros in the real
coefficients case approaches the scaled density of zeros in the complex coefficients case as you move far away from
$\mathbb{R}^m.$

Finally, after working mostly on $\C\backslash\R$ and $\C^m\backslash\R^m,$ we give a weak limit of the error term $\tilde{E}_N(z)$ on compact sets $K \subset \mathbb{C}$ (which may include points in $\R$):
\begin{proposition}[A weak limit, $m=1$]\label{weak1}
\begin{align*}
\frac{1}{N}\tilde{E}_N(z)\,\,dx \wedge dy &= O(N^{-1}), \text{ weakly on compact sets } K\subset \mathbb{C},
\end{align*}
by which we mean that for any $\phi \in C^{\infty} (K),$
\begin{align*}
\frac{1}{N} (\tilde{E}_N(z)\,\,dx \wedge dy, \phi(z)) = \frac{1}{N} \int_K \tilde{E}_N(z) \,\phi(z)\,\,dx \wedge dy  = O(N^{-1}).
\end{align*}
\end{proposition}
\n This means that
\begin{align*}\frac{1}{N}E_{\gamma_{real}}(Z_{f_N}(z)) &= \frac{1}{N}E_{\gamma_{cx}}(Z_{f_N}(z))  +O(N^{-1}) \\&= \frac{1}{\pi(1+|z|^2)^2} + O(N^{-1}),   \text{ weakly on } K \subset \mathbb{C}.
\end{align*}
Note that $K$ could contain some points in $\mathbb{R},$ whereas the strong convergence result excludes points in $\mathbb{R}.$

\subsection{Proof of Lemma \ref{scalederror1} - Scaling limit of the error term}
By the chain rule we have for any differentiable function $f(z)$
\begin{align*}
\frac{\partial^2}{ \partial z \partial \bar{z}} f(z) \Big|_{\frac{z}{\sqrt{N}}} =  N\frac{\partial^2}{ \partial z
\partial \bar{z}} \left[f(\frac{z}{\sqrt{N}})\right]
\end{align*}

So we have from Lemma \ref{eval1}
\begin{align*}
\frac{1}{N} \tilde{E}_N(\frac{z}{\sqrt{N}}) %
&=  \frac{1}{ N\pi }  \frac{\partial^2}{ \partial z \partial \bar{z}}  \log \left[1+2r_N(z)t_N(z)\right]
\Big|_{\frac{z}{\sqrt{N}}}  \\%
&=  \frac{1}{  \pi }  \frac{\partial^2}{ \partial z \partial \bar{z}}  \log
\left[1+2r_N(\frac{z}{\sqrt{N}})t_N(\frac{z}{\sqrt{N}})\right]   ,  z\in\mathbb{C}\backslash\mathbb{R},%
\end{align*}
and after some simplification we get
\begin{align*}
\frac{1}{N} \tilde{E}_N(\frac{z}{\sqrt{N}})  &=  \frac{1}{  \pi }  \frac{\partial^2}{ \partial z \partial \bar{z}}  \log
\left(1+\sqrt{1-\left|\frac{(1+(\frac{z}{\sqrt{N}})^2)^N}{(1+|\frac{z}{\sqrt{N}}|^2)^N}\right|^2}\right)  ,
z\in\mathbb{C}\backslash\mathbb{R}. %
\end{align*}
We now take the limit and get
\begin{align*}
\lim_{N\rightarrow\infty} \frac{1}{N} \tilde{E}_N(\frac{z}{\sqrt{N}}) &= \frac{1}{  \pi } \frac{\partial^2}{ \partial z \partial
\bar{z}}  \log \left(1+\sqrt{1-\left|\frac{e^{z^2}}{e^{|z|^2}}\right|^2}\right)  ,
z\in\mathbb{C}\backslash\mathbb{R}. %
\end{align*}
Setting $z=x+iy,$ we can write
\begin{align*}
\lim_{N\rightarrow\infty} \frac{1}{N} \tilde{E}_N(\frac{z}{\sqrt{N}})&=  \frac{1}{  4\pi } \frac{\partial^2}{ \partial y^2 }  \log
\left(1+\sqrt{1-e^{-4y^2}}\right)  , y \neq 0,  %
\end{align*}
and after simplification and adding to $K^\infty_{\gamma_{cx}}(z) = \frac{1}{\pi},$ we recover Prosen's result in ~\cite{prosen}:
\begin{align*}
K^\infty_{\gamma_{real}}(z)&= \frac{1}{\pi} \,
\frac{1-(4y^2+1)e^{-4y^2}}{(1-e^{-4y^2})^{3/2}}.
\end{align*}

\subsection{Proof of Proposition \ref{scaled1} - Theorem 2 for $m=1$} Using Lemma \ref{scalederror1} we have the asymptotics
\begin{align*}
K^\infty_{\gamma_{real}}(y) = O(|y|)
\end{align*} for $y$ near zero.

\subsection{Proof of Theorem \ref{thm2}}

From the proof of Lemma \ref{exactm} we have
\begin{align} \label{scaledmformula}%
E_{q,\infty}^{\sigma,\tau}(z) :&=    \lim_{N\rightarrow \infty}\frac{1}{N^k} E_{q,N}^{\sigma,\tau} (\frac{z}{\sqrt{N}}) \\&=   \left[   \prod_{\ell \in L_q}   D_\ell^{\sigma,\tau}  ||z||^2    \right]\left[      \prod_{\ell\notin L_q}    D_\ell^{\sigma,\tau}  \frac{1}{2} \log \left(1 + \sqrt{1- \left|\frac{ e^{z \cdot z}}{e^{||z||^2}}\right|^2}\right)    \right]  %
\end{align}
where we use the notation $z \cdot z = z_1^2+ \cdots + z_m^2, ||z||^2 =|z_1|^2+ \cdots + |z_m|^2.$  If we
write $z=x+iy,$ then $z \cdot z = ||x||^2 +2i(x\cdot y) + ||y||^2$.  Since $|e^{2i(x\cdot y)}|=1,$ we can write the second product as
\begin{align*} %
      \prod_{\ell\notin L_q}    D_\ell^{\sigma,\tau}  \frac{1}{2} \log \left(1 + \sqrt{1- e^{-4||y||^2}}\right)    \\ %
\end{align*}
Since the first product is bounded (it is 1 if $\sigma(\ell)=\tau(\ell), $ for all $\ell\in L,$ and zero otherwise),   and the second product goes to zero exponentially fast as $||y||
\rightarrow \infty,$ we have
\begin{align*}
E_{q,\infty}^{\sigma,\tau}(z)\rightarrow 0, \text{ as } ||y||\rightarrow \infty.
\end{align*}
Since this is true for each $i,\sigma,$ and $\tau$ we have
\begin{align*}
\tilde{E}_\infty(z) = \sum_{q=2}^{2^m} \sum_{\sigma, \tau} (-1)^{\sigma+\tau} E_{q,\infty}^{\sigma,\tau}(z) \rightarrow 0, \text{
as } ||y||\rightarrow \infty
\end{align*}

and since $$ K^{\infty}_{\gamma_{real}}(z) = \lim_{N\rightarrow \infty} \frac{1}{N}  E_{\gamma_{real}}(Z_{h_N(\frac{z}{\sqrt{N}})})= \lim_{N\rightarrow \infty} \frac{1}{N} \left(E_{1,N}(\frac{z}{\sqrt{N}}) + \tilde{E}_N(\frac{z}{\sqrt{N}})\right) ,$$ we get
$$ K^{\infty}_{\gamma_{real}}(z) \rightarrow \frac{m}{\pi^m}, \text{ as } ||y||\rightarrow \infty$$

We now look at the behavior of the second term in \ref{scaledmformula} near $\mathbb{R}^m.$  We have the following asymptotics:
\begin{align*}
\frac{1}{2}\log(1+\sqrt{1-\exp^{-4||y||^2}}) &= ||y|| + O(||y||^2), \\
\frac{\partial^2}{\partial y_j \partial y_k} \frac{1}{2}\log(1+\sqrt{1-\exp^{-4||y||^2}}) &= \frac{\partial^2}{\partial y_j \partial y_k}||y|| + O(1)  =O(\frac{1}{||y||}).
\end{align*}
Since in the worst case the second term has $m$ products, we have at worst
\begin{align*} %
      \prod_{\ell\notin L_q}    D_\ell^{\sigma,\tau}  \frac{1}{2} \log \left(1 + \sqrt{1- e^{-4||y||^2}}\right)  =O(\frac{1}{||y||^m}),%
\end{align*}
giving us
\begin{align*}
E_{q,\infty}^{\sigma,\tau}(z) &=O(\frac{1}{||y||^m}) \\
\tilde{E}_\infty(z) &= \sum_{q=2}^{2^m} \sum_{\sigma, \tau} (-1)^{\sigma+\tau} E_{q,\infty}^{\sigma,\tau}(z)  =O(\frac{1}{||y||^m})\\
 &=O(\frac{1}{||y||^m}).
\end{align*}

\subsection{Proof of Proposition \ref{weak1} - Weak limit}

Let $K \subset \mathbb{C}$ be a compact set.  Note that unlike before, we are including points on the real line.  We now show that $\frac{1}{N} \tilde{E}_N(z)\,\,dx \wedge dy$ goes to 0 weakly on $K.$  More specifically, we show that for any $\phi \in C^{\infty} (K),$
\begin{align*}
\frac{1}{N} (\tilde{E}_N(z)\,\,dx \wedge dy, \phi(z)) = \frac{1}{N} \int_K \tilde{E}_N(z) \,\phi(z) \,\,dx \wedge dy = O(N^{-1}).
\end{align*}

Recall that
%\begin{align*}
$\dis \tilde{E}_N(z)\,\,dx \wedge dy  = E(\frac{i}{\pi}\partial \bar{\partial} \log |a \cdot u_N(z)|).$
%\end{align*}
By the definition of the expectation of a distribution, we have
\begin{align*}
(\tilde{E}_N(z)\,\,dx \wedge dy, \phi(z)) &= \left(E(\frac{i}{\pi}\partial \bar{\partial} \log |a \cdot u_N(z)|)\, , \, \phi(z)\right) \\
&= E\left(\frac{i}{\pi}\partial \bar{\partial} \log |a \cdot u_N(z)|\, , \, \phi(z)\right)
\end{align*}
By the definition of the derivative of a distribution, we have
\begin{align*}
 E\left(\frac{i}{\pi}\partial \bar{\partial} \log |a \cdot u_N(z)|\, , \,                                      \phi(z)\right)
=E\left(                                     \log |a \cdot u_N(z)|\, , \, \frac{i}{\pi}\partial \bar{\partial} \phi(z)\right),
\end{align*}
and by the definition of a distribution, we can write this as
\begin{align*}
 E\left(\int_K \log |a \cdot u_N(z)|\,  \frac{i}{\pi}\partial \bar{\partial} \phi(z) \right).
\end{align*}

\n Recall that $E$ denotes expectation with respect to the Gaussian measure $d\gamma_{real}.$ We then have by definition of expected
value that this equals
\begin{align*}
\int_{\mathbb{R^N}} \left(\int_K \log |a \cdot u_N(z)|\,  \frac{i}{\pi}\partial \bar{\partial} \phi(z) \right) d\gamma_{real}(a).
\end{align*}

Since the integrand is bounded, and since $\phi(z)$ does not depend on $a,$ we can switch the order of the integrals and get
\begin{align*}
\int_K \left(\int_{\mathbb{R^N}} \log |a \cdot u_N(z)|\,  d\gamma_{real}(a)\right) \frac{i}{\pi}\partial \bar{\partial} \phi(z).
\,
\end{align*}

Recall that by Lemma \ref{eval1} above we have that the inner integral is $\frac{1}{2} \log (1+2rt),$ so we have
\begin{align*}
\int_K \left(\int_{\mathbb{R^N}} \log |a \cdot u_N(z)|\,  d\gamma_{real}(a)\right) \frac{i}{\pi}\partial \bar{\partial} \phi(z) \,
&= \int_K \frac{i}{2\pi} \log (1+2rt)\, \partial \bar{\partial} \phi(z) \,
\end{align*}

Recall also that $r_N(z)$ and $t_{N}(z)$ are both non-negative by construction, and both are bounded by 1 since $r^2+s^2+t^2=1.$
Both of these conditions are true even on the real line, where $r_N=0$ and $t_N=0$ for all $N.$  This implies the crude estimate
$1 \leq (1+2rt)\leq 3,$ everywhere on $\mathbb{C}$ and, in particular, on $K.$
Since $\phi \in C^{\infty} (K),$ we can write

\begin{align*}
\int_K \frac{i}{2\pi} \log (1+2rt)\, \partial \bar{\partial} \phi(z)
 &\leq \int_K C \frac{i}{\pi}\partial \bar{\partial} \phi(z), \\
&= C ||  \frac{i}{\pi}\partial \bar{\partial} \phi(z) ||_{L^1(K)}
\end{align*}
where $C$ is independent of $N,K,$ and $z,$ including $z$ on the real line, and the $L^1$ norm $||  \frac{i}{\pi}\partial
\bar{\partial} \phi(z) ||_{L^1(K)}$ depends only on $K.$  So then we have that
\begin{align*}
(\tilde{E}_N(z)\,\,dx \wedge dy, \phi(z)) \leq C_K
\end{align*}
where $C_K$ is a constant which depends only on $K.$  We now have want we want:
\begin{align*}
\frac{1}{N} (\tilde{E}_N(z)\,\,dx \wedge dy, \phi(z)) \leq \frac{1}{N} C_K = O(N^{-1}).
\end{align*}

Note that when we consider compact sets $K$ that include part of the real line, the weak limit is the only result we have.  We do not get a strong result because the derivatives of $r,s,t,$ and therefore $\tilde{E}_N(z)$ blow up near the real line.  When we find the weak limit and move the
$\partial\bar{\partial}$ from the $\log$ term to the $\phi$ term as we did above, we avoid this problem:  only the derivatives of
$r$ and $t$ blow up near the real line, not the values of the functions themselves.

%%%%%%%%%%%%%%%%%%%%%%%%%%%%%%%%%%%%%%%%%%%%%%%%%%%%%%%%%%%%%%%%%%%%%%%%%%%%%%%%%
\bibliographystyle{alpha}
\bibliography{thesisbib}

\begin{thebibliography}{BSZ00b}

\bibitem[BBL92]{bbl92}
E.~Bogomolny, O.~Bohigas, and P.~Leboeuf.
\newblock {Distribution of roots of random polynomials}.
\newblock {\em Physical Review Letters}, 68(18):2726--2729, 1992.

\bibitem[BBL96]{bbl96}
E.~Bogomolny, O.~Bohigas, and P.~Leboeuf.
\newblock {Quantum chaotic dynamics and random polynomials}.
\newblock {\em Journal of Statistical Physics}, 85(5):639--679, 1996.

\bibitem[BSZ00a]{bszpl}
P.~Bleher, B.~Shiffman, and S.~Zelditch.
\newblock {Poincar{\'e}-Lelong Approach to Universality and Scaling of
  Correlations Between Zeros}.
\newblock {\em Communications in Mathematical Physics}, 208(3):771--785, 2000.

\bibitem[BSZ00b]{bszuniv}
P.~Bleher, B.~Shiffman, and S.~Zelditch.
\newblock {Universality and scaling of correlations between zeros on complex
  manifolds}.
\newblock {\em Inventiones Mathematicae}, 142(2):351--395, 2000.

\bibitem[DSZ04]{dszcrit}
M.R. Douglas, B.~Shiffman, and S.~Zelditch.
\newblock {Critical Points and Supersymmetric Vacua I}.
\newblock {\em Communications in Mathematical Physics}, 252(1):325--358, 2004.

\bibitem[EK95]{ek}
A.~Edelman and E.~Kostlan.
\newblock {How many zeros of a random polynomial are real?}
\newblock {\em American Mathematical Society}, 32(1):1--37, 1995.

\bibitem[Han96]{hannay}
J.H. Hannay.
\newblock {Chaotic analytic zero points: exact statistics for those of a random
  spin state}.
\newblock {\em J. Phys. A: Math. Gen}, 29(5):L101--L105, 1996.

\bibitem[IZ97]{zeitouni}
I.~Ibragimov and O.~Zeitouni.
\newblock {On Roots of Random Polynomials}.
\newblock {\em Transactions of the American Mathematical Society},
  349(6):2427--2441, 1997.

\bibitem[Kac48]{kac}
M.~Kac.
\newblock {On the Average Number of Real Roots of a Random Algebraic Equation
  (II)}.
\newblock {\em Proceedings of the London Mathematical Society}, 2(6):401, 1948.

\bibitem[Mac09]{bmcrit}
B.~Macdonald.
\newblock {Density of Complex Critical Points of a Real Random Polynomial in
  Several Variables}.
\newblock {\em in final preparation}, 2009.

\bibitem[Pro96]{prosen}
T.~Prosen.
\newblock {Exact statistics of complex zeros for Gaussian random polynomials
  with real coefficients}.
\newblock {\em Journal of Physics A: Mathematical and General},
  29(15):4417--4423, 1996.

\bibitem[Ric54]{rice}
S.O. Rice.
\newblock {Mathematical analysis of random noise}.
\newblock {\em Selected Papers on Noise and Stochastic Processes}, pages
  133--294, 1954.

\bibitem[SV95]{sv95}
L.A. Shepp and R.J. Vanderbei.
\newblock {The Complex Zeros of Random Polynomials}.
\newblock {\em Transactions of the American Mathematical Society},
  347(11):4365--4384, 1995.

\bibitem[SZ99]{SZdist}
B.~Shiffman and S.~Zelditch.
\newblock {Distribution of Zeros of Random and Quantum Chaotic Sections of
  Positive Line Bundles}.
\newblock {\em Communications in Mathematical Physics}, 200(3):661--683, 1999.

\end{thebibliography}
\end{document}